\documentclass[twocolumn,fleqn,10pt]{wlscirep}
\usepackage{graphicx,amsmath,amssymb,color}
\usepackage[utf8]{inputenc}
\usepackage[T1]{fontenc}
\usepackage{bm}
\usepackage{hyperref}
\usepackage{amsmath} 
\usepackage{caption}
\title{The nature of low-temperature spin-freezing in frustrated Kitaev magnets}

\author[1]{U. Jena}
\author[1,2,*]{P. Khuntia}
\affil[1]{Department of Physics, Indian Institute of Technology Madras, Chennai, 600036, India}
\affil[2]{Quantum Centre of Excellence for Diamond and Emergent Materials,
Indian Institute of Technology Madras, Chennai, 600036, India}
\affil[*]{e-mail: pkhuntia@iitm.ac.in}

\begin{abstract}
The subtle interplay between competing degrees of freedom, anisotropy, and spin correlations in frustrated Kitaev quantum materials offers an ideal platform to host non-trivial spin freezing and exotic low-energy excitations. Herein, we elucidate the observed low-temperature spin-freezing as evidenced by thermodynamics, nuclear magnetic resonance (NMR), and inelastic neutron scattering (INS) results in a few representative Kitaev quantum materials adopting Halperin and Saslow and spin jam frameworks. The signature of spin-freezing behavior of these spin-orbit driven frustrated magnets is characterized by a bifurcation of zero-field-cooled and field-cooled magnetic susceptibility below the spin-glass temperature much lower than the characteristic interaction energy scale between spins. The temperature dependence of magnetic specific heat $C_m$ shows a broad
maximum and a lower value of entropy than expected, indicating short-range spin correlations while the $T^2$ dependence of
$C_m$ below the spin-glass temperature suggests gapless excitation spectra in frustrated Kitaev magnets. The field-independent behavior of magnetic specific heat below the freezing temperature implies the presence of exotic low-energy excitations. The aging and memory effect experiments in the Kitaev magnets suggest a non-hierarchical free energy distribution, which differs from the hierarchical organization of conventional spin-freezing. Furthermore, the NMR spin-lattice relaxation rate follows a power law behavior below the spin-freezing temperature, suggesting the persistence of unconventional spin excitation spectra and short range spin correlations that are supported by the inelastic neutron scattering experiments, wherein the imaginary part of the dynamic spin susceptibility is directly proportional to the energy transfer. These distinguishing experimental signatures reflect the emergence of topological spin-freezing in frustrated Kitaev magnets, which can effectively be explained by the Halperin and Saslow (HS) hydrodynamic modes relevant for non-trivial spin glass materials. The linearly dispersive HS modes are hypothesized to account for instigating non-Abelian defect propagation, thereby inducing a spin jam state in the low-temperature regime in frustrated Kitaev magnets. Our investigation demonstrates that HS modes capture the essence of topological spin-freezing, characterized by macroscopic ground state degeneracy, short range spin correlations,  unconventional spin fluctuations,  and linearly dispersive low energy excitations in frustrated two dimensional (2D)  Kitaev magnets with bond-dependent exchange interactions on a honeycomb lattice and its 3D analog hyperhoneycomb that offers a viable ground to extend this framework to a large class of frustrated quantum materials.
\end{abstract}
\begin{document}

\flushbottom
\maketitle

\thispagestyle{empty}

\section*{Introduction}
The synergistic interplay between competing degrees of freedom and anisotropy conspire with frustration-induced quantum fluctuations, offering a viable basis to realize an exotic quantum state with low-energy fractional excitations, promising to address some of the enduring themes in quantum condensed matter~\cite{wen2017colloquium, kitaev2006anyons, khatua2023experimental}. The non-trivial ground states of frustrated quantum magnets at $T\rightarrow 0$, characterized by smooth energy landscapes, can preclude localization in condensed matter~\cite{chandra1993anisotropic}. Quenched disorder such as atomic vacancies or randomness in exchange interactions in frustrated quantum magnets plays a vital role in deforming smooth energy landscapes into rugged energy ones, thereby leading to localization and the emergence of glassiness characterized by slow spin dynamics~\cite{binder1986spin}. 
The emergence of glassiness of topological nature in the low-temperature limit of disorder-free frustrated Mott insulators is attributed to macroscopic ground state degeneracy, unusual spin correlations and exotic low energy excitations~\cite{samarakoon2017scaling}. Landau theory of symmetry breaking is quite successful in describing the conventional phase transitions associated with local order parameters. However, Landau'sLandau's theory is inadequate in capturing the central essence of quantum and topological phases such as quantum
spin liquids, quantum criticality, high-temperature superconductivity, and the
fractional Hall effect that are driven by non-thermal control parameters and interplay between emergent degrees of freedom and electron correlations. The concept of topological order \cite{wen1990topological} was introduced to elucidate the phase transition of such unconventional quantum states under the mathematical formulation of topology. Topologically ordered states are robust against perturbations and are characterized by topological degeneracy, fractional excitations, emergent gauge fields, non-local entanglement, and finite topological entropy at zero temperature that have far-reaching implications in advancing our understanding of fundamental physics and set a stage for the next-generation technologies~\cite{wen1990topological}. 
\begin{figure*}[ht]
\centering
\includegraphics[width=\linewidth]{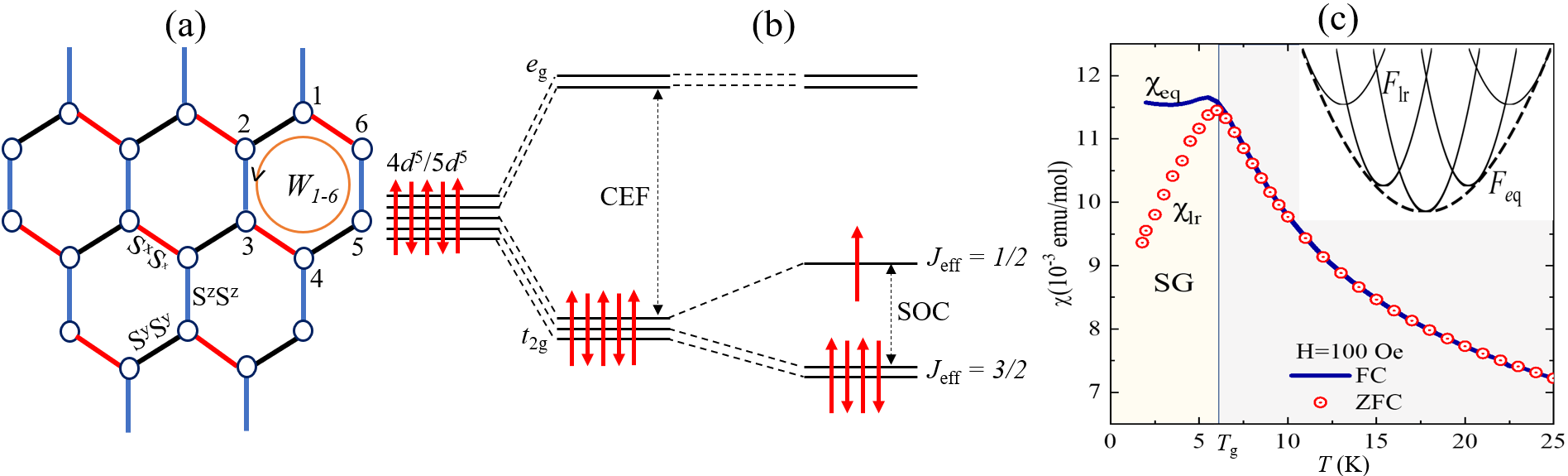}
\caption{(a) A schematic representation of a frustrated Kitaev magnet decorated on a honeycomb lattice with bond-dependent anisotropic exchange interactions (b) The combined effect of crystal electric field (CEF) and spin-orbit coupling in the $4d^5$ and $5d^5$ based quantum magnets realize a low energy $J_\text{eff}=1/2$ state potential to host Kitaev physics (c) The schematics of the temperature dependence of magnetic susceptibility depicting ZFC-FC splitting of magnetic susceptibility below the spin glass temperature in the honeycomb material Li$_2$RhO$_3$ (Adapted from the ref.\cite{khuntia2017local}). The inset shows a schematic representation of a free energy landscape in frustrated spin glass where the solid lines correspond to the possible configuration of free energy in zero field cooling, while the
dashed convex envelop represents the equilibrium free energy
configuration in the presence of magnetic field~\cite{nagaoka2012anderson}. }
\label{fig}
\end{figure*}

A quantum spin liquid state (QSL) is a highly entangled state of quantum matter without a symmetry-breaking phase transition down to absolute zero temperature and is characterized by fractional excitations and long range entanglement~\cite{ anderson1973resonating, khatua2023experimental}. QSL was proposed for $S = 1/2$ moments embodying a Heisenberg triangular lattice \cite{anderson1973resonating}; however, the triangular lattice with nearest-neighbor isotropic Heisenberg exchange interaction host magnetically ordered state. Later, it was demonstrated that the next-nearest neighbor exchange interaction, magnetic anisotropy, and higher-order interactions can stabilize QSL in frustrated magnets. In this context, the spin-orbit driven frustrated honeycomb magnets with bond-dependent highly anisotropic interaction between $J_\text{eff} = 1/2$ degrees of freedom (see Fig. 1a-b) provide a natural habitat to host the celebrated Kitaev QSL state
characterized by a spectrum of deconfined fractional excitations
such as Majorana fermions~\cite{kitaev2006anyons}.
Moreover, the external perturbations, such as magnetic field, break time-reversal symmetry and stabilize a spin gap, supporting the excitation of non-Abelian anyons in the QSL state \cite{kitaev2006anyons, takagi2019concept}. The exactly solvable Kitaev Hamiltonian consists of spin-orbit driven bond-dependent nearest neighbor Ising interactions between $J_\text{eff}=1/2$ moments generated owing to crystal electric field and spin-orbit coupling (see fig. 1a-b) and is represented as $\mathcal{H} = - \sum_{\left<ij\right>} K_{\gamma}S_{i}^{\gamma}S_{j}^{\gamma}$, where $\gamma={x,y,z}$ and $K_{\gamma}$ is the bond dependent Kitaev exchange with the nearest neighbor sites $i$ and $j$ as shown in Fig. 1a \cite{kitaev2006anyons}. The emblematic Kitaev model hosts a solution where the spin-1/2 is fractionalized into four fermions, and alternatively, one can write the Hamiltonian as 
$\mathcal{H} = -\frac{1}{4}\sum_{i,j} K_{\gamma} b_{i}^{\gamma}b_{j}^{\gamma}c_{i}c_{j}$. Subsequently, a set of four Majorana fermions emerges with three immobile fermions \{$b_x, b_y, b_j$\} and one mobile fermion $c$ \cite{takagi2019concept}. The manifestation of Majorana fermions engenders the emergence of a Z$_2$ gauge field. Recently, an emergent glass-like state was observed in the celebrated Kitaev magnet $\alpha$\text{-}RuCl$_3$ in the intermediate magnetic field limit, as evidenced by the non-linear magnetic susceptibility~\cite{holleis2021anomalous} wherein the enhancement of non-linear susceptibility around the spin-glass temperature is related to the fluctuation of magnetic moments in the host \textcolor{blue}{spin-}lattice~\cite{chandra1993anisotropic}. The density-matrix renormalization group method at zero temperature in the intermediate field suggests the emergence of glassiness owing to the slowing down of $Z_2$ fluxes in the proximity of the $U(1)$ spin liquid region \cite{yogendra2023emergent1}. In many-body localized systems, the interplay of many-body interactions and disorder gives rise to an emerging integrability that puts a strong constraint on thermalization~\cite{nandkishore2015many}. 
Recent calculations on one-dimensional spin-1 Kitaev spin chain show the fragmentation of the Hilbert space into unequal disconnected subspaces ~\cite{mohapatra2023pronounced}. In integrable spin systems, the sheer abundance of conserved quantities limits the ability of an initial state to fully navigate all feasible configurations in the Hilbert space and lacks self-thermalization in isolation, leading to the emergence of weak ergodicity breaking in Kitaev materials~\cite{serbyn2021quantum,mohapatra2023pronounced}. In honeycomb Kitaev model, the conserved quantities, for instance, take the form of flux operators defined around each hexagon, expressed as a product of six spin operators defined as $W_{1\text{-}6} = 2^6S_1^zS_2^xS_3^yS_4^zS_5^xS_6^y$ (see Fig. 1a)~\cite{song2016low}. The interplay between the integrability~\cite{kumar2020kitaev} inherent in the Kitaev model and the presence of quenched disorder~\cite{kao2021disorder}, resulting in many-body localization, can lead to the intrinsic glassiness in frustrated honeycomb lattices.

Understanding the origin of spin-freezing in frustrated magnets is of paramount importance, which may provide vital clues for the experimental realization of topological states including the elusive QSL state and associated fractional quantum numbers. Frustration leads to massive ground state degeneracy and low-energy excitations with topological characteristics that are reflected as non-trivial behavior of magnetization, specific heat, NMR relaxation rate, and imaginary part of dynamic spin susceptibility in inelastic neutron scattering in frustrated spin glass in contrast to conventional spin-glass materials~\cite{toulouse1979symmetry}. The experimental search for the topological origin of spin freezing began with the observation of non-trivial spin glass behavior in the magnetopumbites SrCr$_{9p}$Ga$_{12-9p}$O$_{19}$ \cite{ramirez1990strong1} (SCGO) and spinel
Ba$_2$Sn$_2$ZnCr\textsubscript{7p}Ga\textsubscript{10-7p}O$_{22}$ \cite{hagemann2001geometric1} (BSZCGO) decorated on a kagome lattice. In contrast to dilute magnetic alloys, these frustrated magnets feature densely populated magnetic ions potential to exhibit characteristic features that are strikingly different from their canonical counterparts. The manifestation of a glassy state is characterized by the splitting of ZFC-FC magnetic susceptibility (see Fig. 1c) in these systems, which is ascribed to quenched disorder and frustration-induced quantum fluctuations. This unique topological state is commonly referred to as a {`spin jam'} \cite{ yang2015spin, klich2014glassiness}. In such geometrically frustrated systems, the mechanism of {``order-by-disorder''} driven by quantum fluctuations perturbs the classical ground state degeneracy leading to a metastable state within the complex and rugged energy landscape~\cite{chandra1993anisotropic}. 
The spin jam state can be well explained by Halperin-Saslow (HS) mode. Halperin and Saslow proposed a hydrodynamic mode in the background of frozen spins, and a linear dispersion relation is predicted below the freezing temperature \cite{halperin1977hydrodynamic}. HS framework effectively captures the essence of low - temperature spin freezing in the $S=1$ triangular lattice antiferromagnet Ni$_2$Ga$_2$S$_4$ with easy-plane magnetic anisotropy, wherein short-range spin correlations, non-trivial spin fluctuations, and anisotropic spin dynamics govern the low-temperature magnetism~\cite{takeya2008spin, PhysRevB.78.180404, PhysRevB.79.214436}. In 2D frustrated quantum magnets, the $T^2$ behavior of magnetic specific heat ($C_m\propto T^{2}$) can be described by the mean-field low-energy excitations termed as 'spaghetti modes', characterized by a length scale $L_0$ extending upto a order of 10$^2$ number of spins that provide a huge energy barrier to tunnel from one local minima to other ~\cite{klich2014glassiness, podolsky2009halperin, samarakoon2016aging, yang2015spin}. Anderson proposed 
 a scaling behavior where the free energy fluctuation which is the sum of the interaction energy at a generalized plane boundary, separating large blocks of spins, varies as the square root of the boundary area i.e., $\left< \overline{E} \right>\propto A^{1/2}$~\cite{anderson1978concept}. In the spin jam state, finite-length spin folds are present, and the energy fluctuations originate from the interaction between these spin folds, giving rise to metastable states that drive the system to a glassy phase. In this context, the experimental realization and elucidation of unconventional ground states such as spin-glass, associated low-energy excitations and their interactions in spin-orbit driven anisotropic frustrated Kitaev quantum materials decorated on 2D honeycomb and its 3D analog known as hyperhoneycomb lattices may provide key ingredients for establishing paradigmatic models, as well as for the experimental realization, detection, and manipulation of exotic quasi-particles that set a stage for topological quantum computing~\cite{kitaev2006anyons}. However, most frustrated Kitaev magnets exhibit long-range magnetic order due to additional terms in the spin Hamiltonian or stacking faults, which impose a strong constraint on the true experimental realization of the elusive Kitaev spin liquid and exotic Majorana fermions. Similarly, some pristine and doped Kitaev magnets demonstrate spin-glass behavior with an anomalous thermodynamic response, which is strikingly different from that of conventional spin glasses. Nevertheless, there is no detailed analysis to comprehend the nontrivial spin-freezing behavior in frustrated Kitaev magnets. A thorough elucidation of such unconventional spin glass behavior in Kitaev magnets may provide crucial insights for the unambiguous experimental realization of exotic quantum states and quasiparticle excitations in the exactly solvable Kitaev model~\cite{takagi2019concept, jackeli2009mott, yogendra2024fractional}.

  Herein, we provide a comprehensive account demonstrating the unusual spin freezing behavior based on thermodynamic, NMR, and INS results on a complementary scale in a few selected frustrated Kitaev magnets. The low temperature spin freezing phenomenon is interpreted within the framework of the Halperin-Saslow theory, which is relevant for unconventional spin glass behavior, and also incorporates the concept of spin jam. In this framework, the HS modes follow a linear dispersion, and these hydrodynamic modes are entangled to atomic spins that probe the behavior of spins decorated on a frustrated spin lattice at low temperatures.  Our phenomenological interpretation of experimental results in frustrated Kitaev magnets on honeycomb lattice captures the crux of relevant low-energy excitations in the spin glass state. Furthermore, we provide a comparative account concerning the effect of doping on the freezing temperature in conventional spin glass materials and non-trivial spin glass that is observed in frustrated Kitaev magnets. Our analysis and subsequent discussions based on HS hydrodynamic modes reveal that exotic low-energy excitations in frustrated Kitaev magnets are most likely of topological origin, which is entirely different from the conventional spin glass materials.  In essence, the HS formalism has wider applicability in a broad class of frustrated magnets. We also propose a phase diagram taking into account the free energy landscape and observed experimental signatures of Kitaev magnets with distinct magnetic phases, which are highly relevant for the exploration of promising class of frustrated quantum materials for the faithful realization of topological and quantum states with exotic quasi-particle excitations.

\section*{Formalism of Halperin and Saslow (HS) mode}
Halperin and Saslow proposed a nonequilibrium hydrodynamic state for spin glass and helical spin ordering, which completely breaks the $O(3)$ symmetry (locally) \cite{halperin1977hydrodynamic, halperin1969hydrodynamic}. In order to encounter the non-equilibrium states, the magnetization density $m_\alpha(\Vec{r})$ corresponding to three slowly varying rotational angles $\theta_\alpha(\Vec{r})$, $\alpha= 1, 2, 3$ was introduced. The free energy cost because of long-wavelength fluctuation of $\theta_\alpha(\Vec{r})$ and low-frequency fluctuation of $m_\alpha(\Vec{r})$ is expressed by
\begin{equation}
\triangle F[\Vec{m}, \Vec{\theta}] = \frac{1}{2} \sum_{\alpha = 1}^3 \int d^3r\left(\chi^{-1}m_{\alpha}^{2} + \rho_s \arrowvert \Vec{\nabla}\theta_{\alpha}\arrowvert^2\right)
\end{equation}
where $\chi$ and $\rho_s$ are magnetic susceptibility and spin-stiffness, respectively. The spin stiffness characterizes the change in free energy of a frustrated magnet when it undergoes a modulation of spin texture. The spin-texture is identified by short-range spin correlation and oscillating spin density profile owing to topological defects in frustrated magnets. In the spin glass state, the two variables $\theta_{\alpha}$ and $m_{\alpha}$ satisfy the commutation relation
\begin{equation}
   \left[\theta_\alpha(\Vec{r}), m_\beta(\Vec{r'})\right] = i\hslash\gamma \delta_{\alpha\beta} \delta (\Vec{r}-\Vec{r'})
\end{equation}
where $\gamma = g\mu_B/\hslash$ is the gyromagnetic ratio  and $\alpha, \beta$ are the cartesian components correspond to magnetization density along $x,y$ and $z$ directions. 
By applying Heisenberg's equation of motion, which is associated with the commutation bracket stated in equation (2), in conjunction with equation (1), two coupled equations of motion that govern the dynamics of  $m_\alpha$ and $\theta_\alpha$ are represented as \cite{halperin1977hydrodynamic} 
\begin{equation}
    \frac{\partial\theta_\alpha(\Vec{r})}{\partial t}= \gamma\frac{\delta{(\vartriangle F)}}{\delta m_\alpha(\Vec{r})} = \gamma m_\alpha(\Vec{r})\chi^{-1}
\end{equation}
\begin{equation}
    \frac{\partial m_\alpha(\Vec{r})}{\partial t}= -\gamma\frac{\delta{(\vartriangle F)}}{\delta \theta_\alpha(\Vec{r})} = \gamma \rho_s\nabla^2 \theta(\Vec{r})
\end{equation}
By solving the aforementioned pair of coupled differential equations, one can obtain the solutions that describe the dispersion relations for the propagation of low frequency spin wave, which are given by $ \omega = \pm ck$,
where the wave propagation velocity c turns out to be $c = \gamma \left(\rho_s/\chi\right)^{1/2}$. The spin dynamics in densely populated frustrated spin glass materials may instigate two interesting scenarios (i) excitation pertaining to small-amplitude motions around equilibrium or quasi-equilibrium states, known as spin waves, and (ii) barrier mode excitation that involves the large amplitude motions to overcome the energy of metastable state owing to topological defects~\cite{villain1980dynamics}. Gapless spin wave modes can be realized in materials wherein magnetic interactions maintain symmetry under rotations in an n-dimensional spin space~\cite{halperin1969hydrodynamic}. Phenomenologically, below the freezing temperature $T_g$, the Halperin-Saslow (HS) modes exhibit gapless excitation spectrum. The behavior is similar to the observation of sound waves in glasses wherein the fluctuation of random forces is averaged out in the long-wavelength limit~\cite{villain1980dynamics}. Remarkably, the bosonic excitations exhibit distinctive power law characteristics in the magnetic specific heat, namely $C_m \sim T^{D/\mu}$, where $D$ represents the dimension of the spin-lattice and $\mu$ signifies the exponent governing energy dispersion. It should be noted that the observed behavior in magnetic specific heat holds true exclusively for the propagation of undamped modes \cite{fischer1979electrical}. Consequently, in scenarios involving the linear HS mode, the magnetic-specific heat should lead to a $T^2$ behavior in two-dimensional spin-lattice. Similarly, this theory suggests the power-law behavior of nuclear magnetic resonance spin-lattice relaxation rates and linearly dispersive modes, with the imaginary part of the dynamic susceptibility being proportional to the energy transfer in neutron scattering experiments conducted below the spin-glass temperature~\cite{podolsky2009halperin, moriya1956nuclear, hu2021freezing}. This behavior encapsulates the intricate interplay between competing degrees of freedom, anisotropy, macroscopic ground state degeneracy, short range spin correlations, and low-energy excitation, that are manifested in thermodynamic, NMR and INS responses in frustrated magnets, including Kitaev materials.
\begin{figure*}[ht]
\centering
\includegraphics[width=\linewidth]{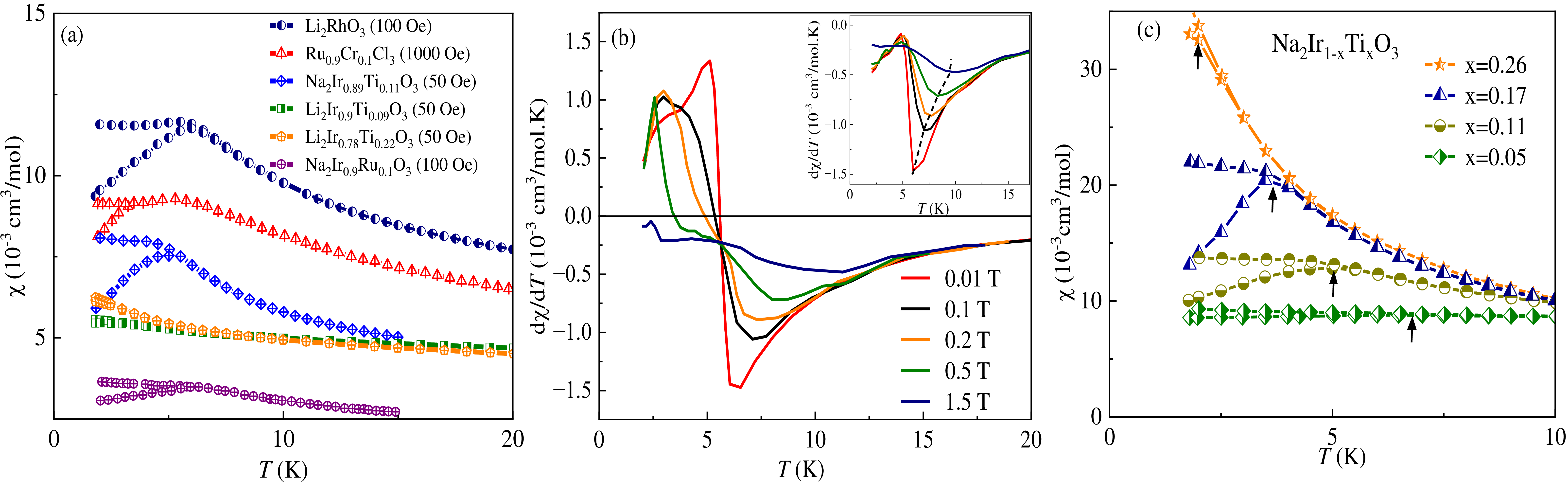}
\caption{(a) The temperature dependence of magnetic susceptibility recorded in zero field and field cooled modes in a few selected frustrated Kitaev magnets decorated on a honeycomb lattice reproduced from the references as cited. The data presented here are adapted from~\cite{khuntia2017local, manni2014effect, bastien2019spin, mehlawat2015fragile}. Magnetic susceptibility of Ru$_{0.9}$Cr$_{0.1}$Cl$_{3}$ and Na$_{2}$Ir$_{0.89}$Ti$_{0.11}$O$_{3}$, scaled by factors of 0.13 and 1.7, respectively, relative to the original data reported in the literature.
(b) The field-dependent first-order derivative of zero-field cooled magnetic susceptibility and the inset feature a field-cooled first-order derivative of magnetic susceptibility of Li$_2$RhO$_3$, reproduced from the reference~\cite{luo2013bingqi}. The arrow marks correspond to the spin-glass transition temperature. (c) The field cooled and zero field cooled susceptibility of Na$_2$Ir$_x$Ti$_{1-x}$O$_3$ for $x = 0.05, 0.11, 0.17$, and 0.26 reproduced from the reference~\cite{manni2014effect} reflecting the suppression of $T_g$ upon increasing doping concentration.}
\label{fig}
\end{figure*}
\section*{Thermodynamics of Kitaev Magnets}

 Competing interaction, spin-orbit driven anisotropy and quantum fluctuations play a vital role for the manifestation of topological quantum states in Kitaev materials. Magnetic susceptibility refers to the response of magnetization in the presence of external magnetic field that provides crucial information concerning the ground state properties of frustrated magnets. In a weak magnetic field, a bifurcation of susceptibility recorded in ZFC and FC modes below the freezing temperature is a characteristic feature of spin glass materials. In ZFC, one can observe a linear response of susceptibility ($\chi_\text{lr}$) due to the fluctuation of magnetization in a particular metastable state characterized by long relaxation time (see Fig. 1c).
 In a given applied magnetic field below the freezing temperature, the spin state exhibits long-lived stability. However, with the gradual increase in temperature, the susceptibility shows an asymptotic enhancement. While in the FC case, as we cool the material from high temperature to low temperature, the system likely transitions to one of the lowest free energy states, representing an average of all ensembles, and the susceptibility can be denoted as $\chi_\text{eq}$ as depicted in Fig. 1c. In terms of the Edward-Anderson spin-glass order parameter, which is associated with the remanent properties of the material under study, the two susceptibilities can be expressed as: $\chi_\text{lr}=\left(\frac{1-q_\text{EA}}{T}\right)$ and $\chi_\text{eq}=\int dq P(q)(1-q)$ \cite{edwards1975theory}.
The spin-glass order parameter \( q_{\text{EA}} \) is defined as the thermal and disorder averaging of the square of the magnetic moment, which can be mathematically expressed as $q_{\text{EA}} = \overline{\left< S_i \right>^2}$~\cite{edwards1975theory}.
The magnetic susceptibility measured in ZFC and FC modes can be represented in terms of free energy as~\cite{fukuyama1982anderson}
\begin{align}
\chi_\text{eq}= \frac{1}{ \left(\frac{\partial^2 F_\text{eq}}{\partial M^2}\right)}
, &\quad
\chi_\text{lr}= \frac{1}{ \left(\frac{\partial^2 F_{lr}}{\partial M^2}\right)}
\end{align}
Here $\chi_\text{eq}$ is the susceptibility corresponds to the free energy at equilibrium ($F_\text{eq}$), which is obtained following the
field cooling protocol and is depicted by the dashed convex envelope in the inset of Fig. 1c. While $\chi_\text{lr}$ is the result of magnetization in the zero field cooled (ZFC) mode, which develops in one of the possible metastable states ($F_{lr}$) enclosed within the quasi-equilibrium free energy state. Since the double derivative is the measure of curvature, it is obvious that $\chi_\text{eq}>\chi_\text{lr}$. This deviation from linear response theory arises due to the extensive exploration of all the phase points in the phase space in field cooling mode. In contrast, under zero field cooling conditions, the system explores only a subset of the phase space, a phenomenon termed as broken ergodicity characterized by long relaxation time compared to the timescale of the experiment~\cite{palmer1982broken}. In the simplest approximation, the difference between the susceptibility measured following two protocols is given by: $\chi_\text{eq}(H)-\chi_\text{lr}(H) = \left(\frac{d^2F_m}{dM^2}\right)^{-1}= \frac{dM_r}{dH}$, where $M_r$ is the remanent magnetization.

Recently, unconventional spin freezing behavior has been observed in Kitaev materials, driven by perturbations such as external magnetic fields~\cite{ holleis2021anomalous, yogendra2023emergent1}, applied pressure~\cite{majumder2018breakdown}, doping~\cite{bastien2019spin, manni2014effect, mehlawat2015fragile}, or quenched disorder~\cite{khuntia2017local}. The frustrated Kitaev magnet Li$_2$RhO$_3$, where 4$d^5$ Rh$^{4+}$ ions constitute a honeycomb spin-lattice with an effective moment of $J_\text{eff} = 1/2$, shows non-trivial spin-freezing with unusual low-energy excitations as reflected in the field magnetic susceptibility, specific heat and NMR relaxation results (Figs. 2a, 3a, 4)~\cite{khuntia2017local, luo2013bingqi, katukuri2015strong}. Notably, in the case of $d^5$ ions in an octahedral environment in 4$d$ and 5$d$ transition metals, the crystal electric field (CEF) splits the energy levels of the $t_{2g}$ and the $e_{g}$ orbitals as shown in Fig. 1b~\cite{ cox2010transition}. Furthermore, the spin-orbit coupling splits the unquenched $t_{2g}$ orbital into Kramer's doublet ground state with $J_\text{eff} = 1/2$ and quartet state with $J_\text{eff} = 3/2$ (see Fig. 1b). For the system with a half-filled Kramer's doublet, the on-site Coulombic interaction induces a gap in the half-filled band, resulting in a weak spin-orbital Mott insulator~\cite{jackeli2009mott}. Experimental observations, in tandem with theoretical calculations, indicate that Li$_2$RhO$_3$ exhibits Mott insulating behavior, characterized by an energy gap $\Delta\sim 80$ meV~\cite{luo2013bingqi}. It shows spin glass behavior around 6 K with remanent magnetization $M_r=1.4\times 10^{-3} \mu_B$/Rh$^{4+}$ at 2 K in the presence of an applied magnetic field $\mu_0 H=100$ Oe~\cite{luo2013bingqi}. 
The first-order derivative of the magnetic susceptibility in Li$_2$RhO$_3$, recorded in both zero-field-cooled (ZFC) and field-cooled (FC) modes, is illustrated in Fig. 2b and its inset, respectively. The inflection point in the ZFC mode, where the slope changes sign (\( \frac{d\chi}{dT} = 0 \)), corresponds to the glass transition temperature in a given magnetic field. This spin freezing behavior is also observed in several 4$d$ and 5$d$ based frustrated Kitaev magnets including Cr doped $\alpha\text{-}$RuCl$_3$, Ti and Ru doped A$_2$IrO$_3$ (A = Na, Li) as presented in this work (Fig. 2), which suggests the existence of a common spin-freezing mechanism in this class of frustrated magnets. The non-magnetic Ti doping at the Ir site in the extensively studied Kitaev magnet Na$_2$IrO$_3$ induces spin freezing and  $T_g$ decreases upon increasing Ti concentration, as shown in  Fig. 2c, which is in stark contrast to that observed in unfrustrated spin-glass materials.

Specific heat experiment is an excellent probe to shed insights into the ground state and associated low-energy quasi-particle excitations in Kitaev magnets. The magnetic specific heat of Li$_2$RhO$_3$ ($C_\text{m}$), obtained after subtracting
the lattice contribution using the specific heat of non-magnetic analog Li$_2$SnO$_3$,
shows $T^2$ dependence (see Fig. 3a) in the low temperature limit below the spin glass temperature suggesting the presence of unconventional low-energy excitations \cite{luo2013bingqi, khuntia2017local}. It may be noted that such $T^2$ dependence is peculiar in the antiferromagnetic two-dimensional Goldstone modes as $C \sim T^D$, where $D$ is the dimensionality of the spin-lattice. In the Kitaev magnet Li$_2$RhO$_3$, the temperature dependence of magnetic specific heat is robust against the external magnetic field as shown in the inset of Fig. 3a, measured up to 9 T. The $C_\text{m}(H, T)/T$ vs. $T$~\cite{khuntia2017local} shows the change of curvature at about the broad maximum $T_\text{broad} = 10$ K, which ensures that all field derivative of $C_\text{m}(H, T)/T$ vanishes at $T_\text{broad}$. Considering $\partial^2C_ \text{m}/\partial^2H = \partial^2\chi/\partial^2T$, it is expected that the first-order temperature derivative of the field-cooled magnetic susceptibility exhibits an extremum around $T_\text{broad}$~\cite{ramirez2000entropy}.  However, it is the minima observed in $d\chi/dT$ as depicted in the inset of Fig. 2b (dashed line) for all magnetic fields up to 1.5 T occur below $T_\text{broad}$, derived from the specific heat data. Since specific heat is more sensitive to low-energy hydrodynamic modes and short range spin correlations, this attribute might provide a plausible explanation for the disparate features observed in the two measurements or different experimental techniques with varied characteristic time scales probe a bit different spin dynamics. This scenario is in sharp contrast to that observed in conventional spin glass, where the system cannot allocate enough time to respond to the dynamically evolving magnetic field and temperature fluctuations in specific heat measurements and is less sensitive to glassiness. Furthermore, another reason behind the disparity could be attributed to the distinct energy scales involved~\cite{syzranov2022eminuscent}. The energy scale $T_\text{broad}$ refers to a hidden energy scale in geometrically frustrated systems associated with the development of short-range correlations among atomic spins as the temperature decreases. It is considered an inherent property of the pure compound and is typically an order of magnitude smaller than the Weiss constant $\theta_{CW}$. The significance of $T_\text{broad}$ lies in its suggestion that a glass transition may occur even in a disorder-free system, indicating a fundamental characteristic of the material's magnetic behavior~\cite{syzranov2022eminuscent}.

In Li$_2$RhO$_3$, the magnetic entropy recovered at 45 K is much lower ($0.35 \text{R} \ln 2$) than that expected for $J_{\text{eff}}=1/2$ moments suggesting huge ground state degeneracy and short-range spin correlations in agreement with the presence of a broad maximum in specific heat and NMR spin-lattice relaxation rate \cite{khuntia2017local}. The specific heat rules out the presence of any Schottky contribution, and the observed behavior of magnetic specific heat in Li$_2$RhO$_3$ is at variance with that found in conventional spin glass materials wherein $C_m\sim T$~\cite{anderson1972anomalous}.  Remarkably, similar  $C_m\sim T^2$  behavior has been observed in several Kitaev materials such as Cr doped $\alpha\text{-}$RuCl$_3$~\cite{bastien2019spin}, Ti doped Li$_2$IrO$_3$~\cite{ manni2014effect},  Ru doped Na$_2$IrO$_3$~\cite{mehlawat2015fragile} below the spin-freezing temperature much below the characteristic exchange interaction as shown in Fig. 3a. To shed insights into this unconventional low-energy excitations as reflected in the specific heat of the spin-orbit driven frustrated Kitaev magnets with a goal to demonstrate a commonality in this class of materials, we have analyzed the specific heat of a few representative Kitaev quantum magnets following Halperin-Saslow formalism. The coherent propagation of HS mode can be captured with a finite spin-stiffness constant in the spin texture with a characteristic length scale $L_0$. Within this finite spin-texture, the low-energy spin excitation in $2D$ spin-lattice gives rise to the magnetic specific heat 
$\frac{C_m}{R} = \left[\frac{3\sqrt{3}\zeta(3)}{2\pi}\right]\left(\frac{ak_BT}{\hslash \mathcal{D}}\right)^2 - \left(\frac{\sqrt{3}\pi}{2}\right)\left(\frac{a}{L_0}\right)^2 $ at temperature $T<T_g<<\arrowvert \theta_{CW}\arrowvert$, where $\theta_{CW}$ refers to the energy scale of exchange interaction and $\mathcal{D}$ is the spin stiffness constant associated with the free-energy~\cite{nakatsuji2005spin2, ramirez1992elementary}. It is worth noting that we employ the spin stiffness constant $\mathcal{D}$ for dimensional normalization, distinct from the notation $\rho_s$ used previously for the same quantity and $L_0$ is the characteristic length scale over which the Goldstone modes are well defined. The first term corresponds to the quadratic temperature dependency of $C_\text{m}$ in $2D$ antiferromagnetic spin wave, and the second term is the size-dependent negative shifting of specific heat valid for a $2D$ gapless linear dispersive mode with frequency $\frac{\hslash\omega}{k_B} <T< \arrowvert \theta_{CW}\arrowvert$ \cite{ramirez1992elementary}. Considering only the antiferromagnetic ordering at $T\sim \arrowvert \theta_{CW}\arrowvert$, the spin spin stiffness constant can be calculated from the relation $\mathcal{D}_{0}^2= \left[3\sqrt{3}\zeta(3)/4\pi\right] (ak_B\theta_{CW}/\hslash)^2/\ln(2S+1)$ \cite{nakatsuji2005spin2}. The fit of the low temperature experimental specific heat data below  $T_g$ of Li$_2$RhO$_3$ with the expression relevant for elucidating the low-energy excitations in the HS mode yields, $\mathcal{D}_0 \sim 1960$ m/s and $\mathcal{D} \sim 1091$ m/s with spin texture length $L_0 \sim 40$ nm  and corresponding fit is shown in Fig. 3b. This framework has been extended to Cr doped $\alpha\text{-}$RuCl$_3$, Ti doped Li$_2$IrO$_3$,  Ru doped Na$_2$IrO$_3$  Kitaev materials and the resulting parameters are presented in Table 1 exemplifying non-trivial spin-freezing mechanism in this class of spin-orbit driven frustrated quantum materials. The reduction in stiffness constant $\mathcal{D}$ compared with $\mathcal{D}_0$ is ascribed to softening owing to magnetic frustration~\cite{podolsky2009halperin}.\\ 
\begin{figure*}[ht]
\centering
\includegraphics[width=\linewidth]{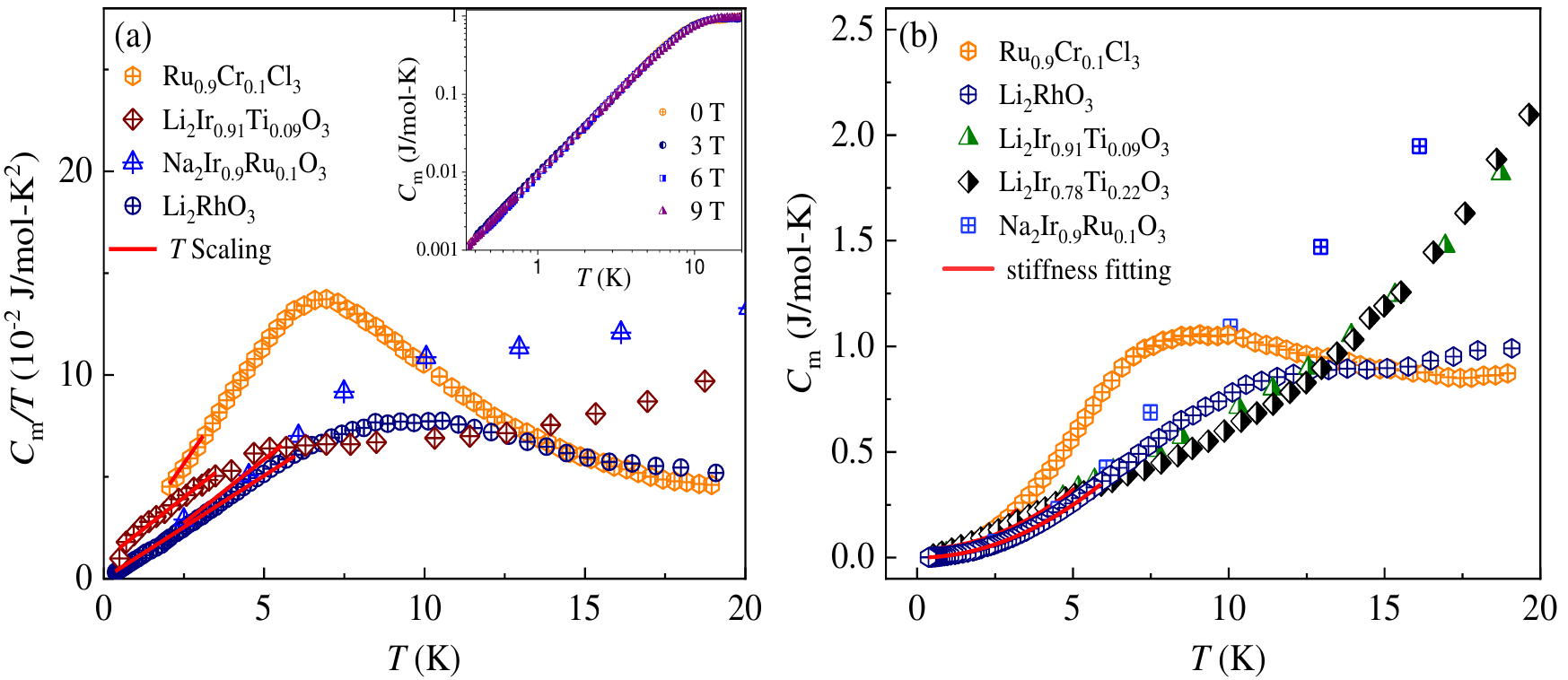}
\caption{(a) The temperature dependence of magnetic specific heat in selected Kitaev materials and the red solid line depicts a power law fit ($C_\text{m}\sim T^2$) below the spin-glass temperature. The inset shows the field independent behavior of magnetic specific heat in Li$_2$RhO$_3$ on a double log-scale. By examining the trend of $C_\text{m}/T$ as a function of temperature, we aim to elucidate the unique thermodynamic properties and underlying physics that govern these Kitaev magnets (b) The temperature dependence of magnetic specific heat of Kitaev magnet with spin stiffness fitting below $T_g$ as discussed in the text, offering insights the ground state properties. The data presented here are adapted from~\cite{khuntia2017local, manni2014effect, bastien2019spin, mehlawat2015fragile}.}
\label{fig}
\end{figure*}
\begin{table*}[ht]
\centering
\begin{tabular}{|l|l|l|l|l|l|l|l|}
\hline
Kitaev Magnet  & $T_g$ & $\arrowvert \theta_{CW}\arrowvert$ (K)& $f=\frac{\arrowvert \theta_{CW}\arrowvert}{T_g}$ & $\tau_1$(K) & $\tau_2$(K) & $\mathcal{D}_0$(m/s) & $\mathcal{D}$(m/s) \\
\hline
Li$_2$RhO$_3$~\cite{khuntia2017local} & 6 &60&10 & 80 & 19 & 1960 & 1090 \\
\hline
Ru$_{0.9}$Cr$_{0.1}$Cl$_3$~\cite{bastien2019spin} & 3.1 & 85&28&-&-&3890&878\\
\hline
Na$_2$Ir$_{0.89}$Ti$_{0.11}$O$_3$~\cite{manni2014effect} & 5 & 55&11&250&4.7&1926&-\\
\hline
Na$_2$Ir$_{0.9}$Ru$_{0.1}$O$_3$~\cite{mehlawat2015fragile} & 5.6 &127&23&178&5.5&4300&1080\\
\hline
Li$_2$Ir$_{0.91}$Ti$_{0.09}$O$_3$~\cite{manni2014effect} & 3.5 & 33&9&135&8&1095&900\\
\hline
Li$_2$Ir$_{0.78}$Ti$_{0.22}$O$_3$~\cite{manni2014effect} & 2 & 33&16&120&10&1095&690\\
\hline
\end{tabular}
\caption{\label{tab}Depicts the relevant parameters extracted from the analysis of magnetic susceptibility and specific heat following Halperin-Saslow framework for a few selected frustrated honeycomb Kitaev magnets exhibiting non-trivial spin-freezing behavior where the degree of frustration is quantified by the frustration parameter $f=\frac{\arrowvert \theta_{CW}\arrowvert}{T_g}$ in these magnets.}
\end{table*}
\vspace{-\baselineskip}

Next, the $T^2$ dependence of magnetic-specific heat, as shown in Fig. 3a in the Kitaev magnets wherein the magnetic frustration is mediated by spin-orbit driven anisotropic bond-dependent exchange interaction, can be explained by assuming three non-degenerate hydrodynamic modes in the presence of the magnetic field. With the application of magnetic field $H$, the degeneracy of hydrodynamic mode gaps out, and the frequency of each mode is expressed as \cite{podolsky2009halperin}:$\omega_{\pm} = \pm \frac{g\mu_BH}{2}+\sqrt{\left(\frac{g\mu_BH}{2}\right)^2+(ck)^2}$ where $g$ is the  Landé $g$-factor, $c$ is the velocity of hydrodynamic modes and $k$ is the wave vector. In the limit of $\left(\frac{g\mu_BH}{2}\right)>> ck$, the Taylor expansion yields $\omega_+ \approx g\mu_BH + \frac{c^2k^2}{g\mu_BH}$ and  $\omega_- \approx \frac{c^2k^2}{g\mu_BH}$. While $\omega_0 = ck$ corresponds to field-independent excitations showing linear dispersion relation. The quadratic polarization $\omega_-$ compensates for the gapped mode $\omega_+$ to some extent at low energy. This compensation accounts for the observed negligible deviation from the $T^2$ behavior of $C_m$ at low temperature below the spin-glass temperature. The proportionality constant A ($C_m = AT^2$) is given by: $A = \frac{3\zeta(3)k_B^2RV}{\pi d \hslash^2}\sum_i \frac{1}{c_i}$, where $V$ is volume of the unit cell, $d$ is the spacing between the successive layers in the honeycomb lattice, and $c_i$ is the velocity of spin wave mode along three spatial directions. Apart from the HS mode, the robustness of specific heat against an external magnetic field below the freezing temperature, can be associated with pseudo-Goldstone modes fostered by noncollinear antiferromagnetic ordering ~\cite{kawamura2007vortex}. An applied external magnetic field reduces the symmetry of the Hamiltonian from \( O(3) \) to \( O(2) \), decreasing the number of Goldstone modes from three to one. Despite this symmetry reduction, the noncollinear AF ground state retains an "accidental" degeneracy, unrelated to the \( O(2) \) symmetry of the Hamiltonian, similar to that in zero-field conditions~\cite{kawamura1985phase}. The ground-state manifold preserves three adjustable continuous parameters, including one true Goldstone mode from symmetry and two pseudo-Goldstone modes not linked to the Hamiltonian symmetry. These pseudo-Goldstone modes play a crucial role in maintaining the \( T^2 \) behavior of the low-temperature specific heat under applied magnetic fields~\cite{kawamura1985phase, kawamura2007vortex}. Thus, the specific heat remains robust against external perturbations. Noteworthy accidental degeneracy and pseudo-Goldstone modes may become approximate in quantum systems due to quantum effects~\cite{kawamura2007vortex}.
A broad peak in $C_m/T$ is observed around temperature $T = 10$ K, indicating the persistence of short-range spin correlations in Li$_2$RhO$_3$~\cite{khuntia2017local}. In an isotropic unfrustrated system, a $\lambda$-type peak in specific heat is typically associated with the onset of long-range magnetic ordering which can be suppressed completely by applying a critical field $H_C(0)\sim k_B\arrowvert \theta_{CW}\arrowvert/g\mu_B$. So one would expect a complete suppression of the peak in specific heat by applying a critical magnetic field of approximately $H_C(0)\sim k_BT_\text{peak}/g\mu_B = 7.4$ T if the peak at $T=10$ K was associated with long range magnetic ordering in Li$_2$RhO$_3$ \cite{ khuntia2017local}. However, it is worth noting that, even in an applied field of 9 T, there is no suppression of the observed peak in Li$_2$RhO$_3$. A similar scenario is also observed in Cr$\text{-}$doped $\alpha\text{-}$RuCl$_3$, Ti doped Li$_2$IrO$_3$,  Ru doped Na$_2$IrO$_3$. The broad peak in specific heat and $T^2$ behavior of $C_\text{m}$ below the spin-glass temperature (see fig. 3) are robust against external magnetic field, which is associated with the presence of short-range spin correlations and abundant low energy excitations in these frustrated magnets~\cite{bastien2019spin, manni2014effect, mehlawat2015fragile}. Second, it can be the coexistence of glassiness with spin singlets formation around the energy scale of glass transition $T_\text{g}$~\cite{ramirez2000entropy}. The low-energy state might be dominated by these moment free singlets. It may represent a specific type of spin correlation characterized by a Gaussian-shaped relaxation profile in muon spin resonance ($\mu$SR) experiment involving short-lived singlets whose life-time is significantly shorter than that of the muon~\cite{vzivkovic2021magnetic, ramirez2000entropy}. For instance, in the frustrated trillium spin-lattice K$_2$Ni$_2$(SO$_4$)$_3$, this scenario involves proximity to a quantum critical point (QCP) between an ordered phase and a quantum fluctuation-dominant phase, with K$_2$Ni$_2$(SO$_4$)$_3$ near the ordered side~\cite{vzivkovic2021magnetic}. It escapes the glassiness region; however, another trillium lattice, KSrFe$_2$(PO$_4$)$_3$ with high spin (S = 5/2), exhibits the emergence of glassiness characterized by a power-law specific heat $C_\text{m}\sim T^{2.33}$ below spin-glass temperature~\cite{boya2022signatures}. The trillium lattice material KSrFe$_2$(PO$_4$)$_3$ that crystallizes in the cubic space group forms a diamond spin-lattice through the combination of nearest-neighbor \(J_1=-4.87\) K and next-nearest-neighbor \(J_2=-2.67\) K couplings. The power law behavior of specific heat $C_\text{m}\sim T^{2.33}$  in diamond lattice antiferromagnets arises from the quantum treatment of thermal fluctuations, particularly when considering the behavior of the fluctuation stiffness parameter \(\kappa_T(q)\) at low temperatures~\cite{bergman2007order}. In these systems, \(\kappa_T(q)\) is modified non-perturbatively by thermal fluctuations, leading to a non-analytic temperature-independent factor that vanishes only at specific spiral wavevectors \(\pm Q\), corresponding to the emergence of Goldstone modes. This results in a non-analytic temperature dependence in the specific heat. While the classical specific heat scales as \(C_{\text{classical}}(T) = A + BT^{1/3}\) at low temperatures, the quantum treatment, which involves quantization of the spin-wave modes to obtain the magnon spectrum, predicts a specific heat power law with an exponent of \(7/3\): \(C_{\text{quantum}}(T) \sim T^{7/3}\) ~\cite{bergman2007order}. This behavior reflects the intricate interplay between thermal fluctuations, the fluctuation stiffness \(\kappa_T(q)\), and the non-analytic temperature dependence, highlighting the unique quantum behavior and thermal properties of the spin system under consideration.

In the HS framework, the magnetic susceptibility $\chi_m$, defined as the experimental value of the ZFC susceptibility as the temperature approaches zero, and $C_m/T^2$ below $T_g$, can be linked to yield two characteristic energy scales denoted as $E_1$ and $E_2$, which are expressed as follows:
\begin{equation}
    E_1 = \frac{2g^2\mu_{B}^2S(S+1)N_A}{z\chi_m} = k_B \tau_1
\end{equation}	
The energy scale \(E_2\) is derived from the relation~\cite{halperin1977hydrodynamic, podolsky2009halperin}:
\[
\frac{C_m}{N_A \nu} = \frac{3 \zeta(3)}{\pi d} \frac{k_B^3 T^2}{\hslash^2} \sum_i \left( \frac{1}{c_i^2}\right)  - \frac{3 k_B \pi}{L_0^2 d} \tag{3}
\]
where \(C_m\) is the magnetic specific heat, \(\nu = \frac{\sqrt{3}a^2 d}{2}\) is the unit cell volume, \(d\) is the spacing between decoupled spin-lattice layers, \(a\) is the bond length between magnetic moments in a plane, and \(L_0\) is the characteristic length scale for Goldstone modes. Neglecting the second term and considering three Goldstone modes (\(n_p = 3\)):

\[
\frac{C_m}{N_A \nu} = \frac{3 \zeta(3)}{\pi d} \frac{k_B^3 T^2}{\hslash^2} \frac{3}{c^2} \tag{4}
\]

Using \(c = \gamma \left( \frac{\rho_s}{\chi_m} \right)^{1/2}\) in equation (4):

\[
\frac{\rho_s}{N_A \nu} = \frac{3 n_p \zeta(3)}{\pi d} \frac{k_B^3}{g^2 \mu_B^2} \frac{\chi_m}{C_m / T^2}
\]

For in-plane interactions, this gives the energy scale \(E_2\):

\[
E_2 = \frac{\rho_s d}{N_A \nu} = \frac{3 n_p \zeta(3)}{\pi} \frac{k_B^3}{g^2 \mu_B^2} \frac{\chi_m}{C_m / T^2}
\]
\begin{equation}
    E_2 = \frac{3n_p\zeta(3)}{\pi}\frac{k_B^3}{g^2\mu_B^2}\frac{\chi_m}{C_m/T^2} = k_B\tau_2
\end{equation}
Here $z$ is the co-ordination number i.e., the number of nearest neighbor magnetic ions, which is 3 for the honeycomb lattice, and $n_p$ is the number of degenerate hydrodynamic modes. The two energy scales $E_1$ and $E_2$, are linked to the exchange interaction energy and spin stiffness, which are in turn comparable to two distinct temperature scales, $\arrowvert \theta_{CW}\arrowvert$ and $T_g$, respectively. The external perturbations, such as doping at the magnetic site of Kitaev magnets, could lead to novel physical phenomena such as spin freezing in the case of Ru$_{1-x}$Cr$_x$Cl$_3$ or Na$_2$Ir$_{1-x}$Ru$_x$O$_3$. In such a scenario, the relevant energy scale can be extracted following modified expressions relevant for $E_1$ and $E_2$.
The derived relevant energy scales for some selected Kitaev quantum materials are tabulated in Table-1, which suggests a common underlying mechanism governing unconventional spin-freezing phenomena in frustrated Kitaev magnets.

In quantum materials with ferromagnetic interactions, the undamped linear dispersion of spin wave mode leads to a $T^3$ behavior in specific heat
  below the spin glass temperature \cite{fischer1979electrical}. Such behavior is observed in the 3D variant of Kitaev honeycomb magnet, namely, the hyperhoneycomb lattice $\beta\text{-}$ZnIrO$_3$ and $\beta\text{-}$MgIrO$_3$, where the dominant exchange interaction between effective spin $J_\text{eff} = 1/2$ of Ir$^{4+}$ moments is ferromagnetic with $\theta_{CW}\sim 45.6$ K and 56 K, respectively \cite{haraguchi2022quantum, zubtsovskii2022topotactic, haraguchi2023monoclinic}. In this hyperhoneycomb spin-lattice, the Z$_2$ flux operator $W$, is a conserved quantity defined as a loop operator that encompasses precisely ten lattice sites. Both 3D Kitaev materials, $\beta$-ZnIrO$_3$ and $\beta$-MgIrO$_3$, demonstrate a distinctive spin freezing phenomenon, characterized by the onset of weak anomaly in thermodynamic experiments at temperatures around 12 K and 22 K, respectively~\cite{haraguchi2022quantum, zubtsovskii2022topotactic,  haraguchi2023monoclinic}.
The magnetic contribution to the specific heat exhibits a notable $T^3$ dependency below the spin-glass temperature, indicating a linear dispersion behavior, which was initially postulated to elucidate the hydrodynamic mode of spin waves in ferromagnetic materials~\cite{halperin1969hydrodynamic}. For ferromagnetic long-range ordering, the spin wave excitations lead to quadratic dispersion relation so the linear dispersion is due to planar ferromagnetic excitations.
The $T^3$ behavior remains intact for $\beta\text{-}$ZnIrO$_3$ in the presence of an applied magnetic field upto 5 T, which is due to the compensation of $\omega_+$ with $\omega_-$ and the propagation of linear mode ($\omega_0$) accounts for the field independent behavior of specific heat in the presence of an external magnetic field \cite{haraguchi2022quantum}.

Next, we demonstrate how the Kitaev spin freezing of topological origin manifested by macroscopic ground state degeneracy and linearly dispersive low energy spin excitations, significantly differs from conventional spin-glass materials. In a seminal paper \cite{anderson1972anomalous}, Anderson \textit{et al.} introduced the idea of two-level system (TL), which successfully describes the linear behavior of specific heat in conventional spin glass. At low temperatures, only tunneling systems with energy differences \( \Delta E \) close to \( k_B T \) contribute significantly to the specific heat. The specific heat \( C \) scales linearly with temperature \( T \) because the density of states for \( \Delta E \) is approximately constant around zero energy difference. This behavior aligns with the Anderson-Varma-Halperin (AVH) mechanism, which requires the presence of localized tunneling states. For tunneling to occur, the energy barrier separating two local minima must not be too large, enabling transitions between states that are energetically degenerate within \( k_B T \). The specific heat contribution from a cluster spin glass, incorporating these tunneling effects, can be expressed as $
C_{TL} = \frac{\pi^2}{6}k_B^2 T n(0),$
where \( n(0) \) represents the density of states\cite{anderson1972anomalous}. This density of states accounts for the linear specific heat behavior, with the energy scale \( \Delta E \sim k_B T_g \). The absence of the AVH mechanism in 2D Kitaev magnets, despite spin freezing, stems from their quantum and topologically constrained nature. The AVH mechanism relies on a broad, random distribution of local magnetic fields (\(H_{\text{loc}}\)) with a nonzero density of states at \(H_{\text{loc}} = 0\), enabling linear specific heat via localized zero-point tunneling~\cite{PhysRev.132.2412, PhysRev.118.1519}. In contrast, the Kitaev Hamiltonian \(H = -\sum_{\langle i,j \rangle} J_\gamma S_i^\gamma S_j^\gamma\) imposes anisotropic, bond-dependent interactions that create a structured energy landscape dominated by emergent flux excitations and gapless excitations~\cite{kitaev2006anyons, PhysRevB.78.115116}. Enhanced quantum fluctuations in 2D suppress classical trapping, while topological constraints preclude the singular field distribution at \(H_{\text{loc}} = 0\)~\cite{hermanns2018physics}. Thus, Kitaev magnets epitomize a fundamentally different paradigm, where spin freezing emerges from quantum coherence and geometrical frustration, not the random field physics central to AVH. In principle, the contribution to the specific heat arises from both spin jam and spin glass cluster mechanisms~\cite{piyakulworawat2023zero}. As such, the specific heat can be expressed as a linear combination: $C = f \cdot C_{HS} + (1-f) \cdot C_{TL}$, where $C_{HS}$ represents the specific heat contribution from the spin jam component, while $C_{TL}$ is that due to the spin glass cluster, and $f$ is a weighting factor corresponds to the relative proportion of each contribution. The magnetic susceptibility and specific heat provide a consistent picture in accord with the HS framework manifesting a common spin-freezing mechanism in a few selected Kitaev magnets presented here.

\section*{Local Probe Techniques}
Nuclear magnetic resonance and neutron scattering are two excellent microscopic techniques to shed insights into non-trivial spin glass characterized by short-range spin correlations and unconventional low energy spin excitations in frustrated Kiatev magnets. The NMR spectra broaden upon lowering temperature below the spin glass temperature. 
\begin{figure}[htpb]
		\begin{center}
			\includegraphics[width=0.45\textwidth,height=0.38\textwidth]{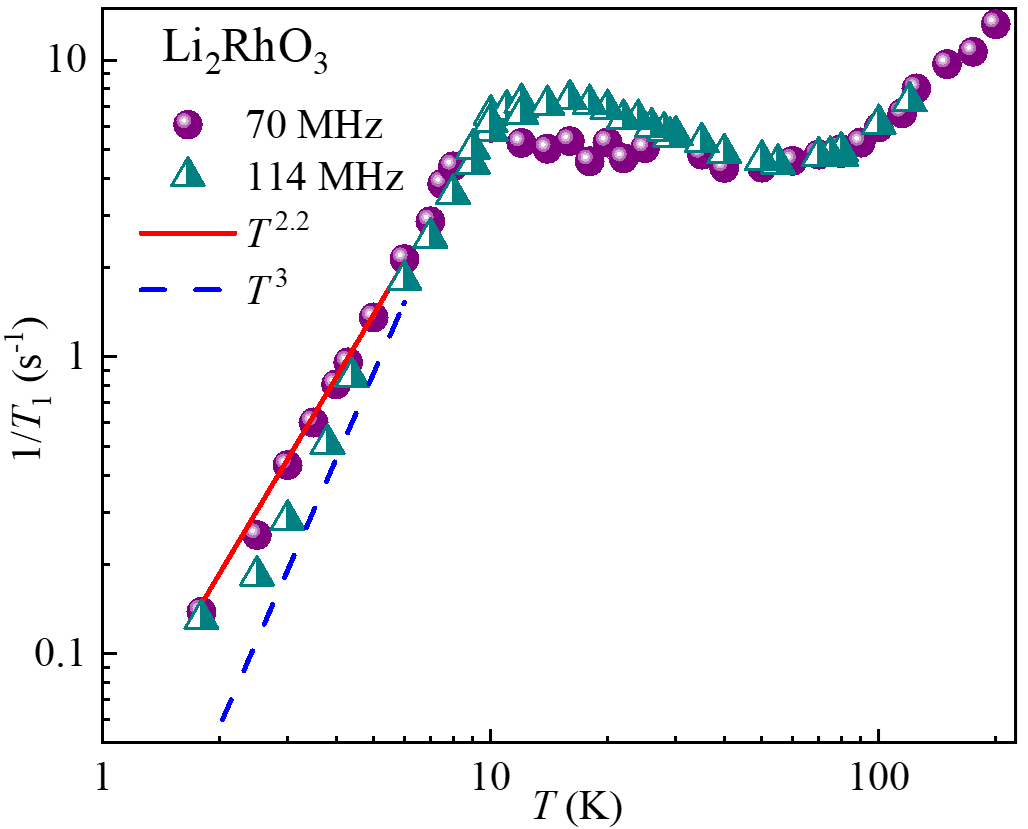}
			\caption{The temperature dependence of NMR relaxation rate \(1/T_1\) in Li\(_2\)RhO\(_3\) taken in two different magnetic fields show a \(T^{2.2}\) dependence (solid red line). The
data presented here is adapted from~\cite{khuntia2017local}}
			\label{fig: energies_pp_bar42111_4_13}
		\end{center}
	\end{figure}
The relative NMR linewidth defined as the linewidth divided by the resonance field (\(\delta H = \Delta H / H\)) at two NMR frequencies 70 MHz ($\mu_0H\sim 4.23 T$) and 114 MHz ($\mu_0H\sim 6.89 T$) exhibits no field dependency below the spin-glass temperature (Fig. 4a of Ref.~\cite{khuntia2017local}). This transition is marked by a gradual development of frozen moments, where intermediate temperatures display both static and dynamic components. In general, the magnetic broadening in $\Delta H$ reflects the distribution of internal magnetic fields. In paramagnetic systems, these fields originate from induced moments that vary linearly with the applied magnetic field $H$. However, during the transition to a frozen state with a local static moment, the width $\Delta H$ is expected to remain largely unaffected by changes in $H$ that is consistent with magnetic susceptibility~\cite{quilliam2011ground}. The NMR line width is independent of temperature in the paramagnetic region  $T>>\theta_{CW}$, however, the enhancement of NMR line width in the intermediate temperature range well above the spin glass transition suggests the predominance of short-range spin correlations that are in agreement with thermodynamic results. NMR spin-lattice relaxation rate $1/T_1$,  that tracks low-energy spectra in frustrated magnets follows $T^{2.2}$ (Fig. 4) behavior down to 1.8 K in Li$_2$RhO$_3$ suggesting the persistence of low energy excitations below spin-glass temperature $T_g \sim 6$ K. The deviation of the exponent of the power law in the temperature dependence of $1/T_1$ from the expected $1/T_1\sim T^3$ behavior might be related to the distribution of local hyperfine fields at the probing nuclear site reflected as broad NMR spectra and/or anisotropic spin fluctuations at low temperatures~\cite{khuntia2017local, moriya1956nuclear, takeya2008spin}.
The broad maximum around 10 K in $1/T_1$, well above \( T_g \), infers the persistence of short-range spin correlations, which is consistent with the broad peak in the magnetic specific heat. The absence of a significant loss in NMR signal intensity and the power law behavior of NMR spin-lattice relaxation rate further support that Li\(_2\)RhO\(_3\) is not a conventional spin-glass material. The spin-lattice relaxation rate, \( \frac{1}{T_1} \), is related to the spin correlation as the wave vector averaged dynamical susceptibility \(\chi_m^{\prime\prime} (\mathbf{q},\omega)\), 
\(\frac{1}{T_1} \propto \sum_{\mathbf{q}} |A_\text{hf}(\mathbf{q})|^2 \frac{\chi_m''(\mathbf{q}, \omega)}{\omega}\), which probes low energy spin excitations in frustrated magnets. For instance, the frustrated triangular lattice antiferromagnet NiGa$_2$S$_4$ exhibits an anisotropic spin dynamics and power-law $T^3$ dependence below 1 K in NMR NMR spin-lattice relaxation rate, $1/T_1$ \cite{takeya2008spin}. The linear dispersing mode in 2D magnets results in a specific heat exhibiting a quadratic temperature dependence ($C_m \propto T^2 N(E_F)$), where $N(E_F)$ is the density of states at the Fermi level, and the spin-lattice relaxation rate follows a cubic temperature dependence ($1/T_1 \propto T^3 N^2(E_F)$).
 The damping of spin wave excitation gives rise to a dynamic susceptibility of the form $\chi^{\prime\prime} (k,\omega) = \frac{\omega\chi D_sk^2}{2}\left[\frac{1}{(\omega-ck)^2+(D_sk^2)^2}+\frac{1}{(\omega+ck)^2+(D_sk^2)^2}\right]$ with $D_s$ is the spin diffusion constant \cite{fischer1979electrical}. In the topologically protected state, the barrier mode propagation involves the diffusion of quadruplet, not pair, flips of defect variables by spin operators. So, the barrier mode is thermally suppressed, and in the limit of $D_s\rightarrow 0$, the imaginary part of the dynamic susceptibility becomes $\chi^{\prime\prime}= \frac{\pi}{2}\omega\chi\left[\delta(\omega-ck)+\delta(\omega+ck)\right]$. This results in the propagation of undamped spin wave modes that, in conjunction with linear modes, yield a power-law dependence in specific heat, which reflects the gapless nature of the low-energy excitations~\cite{fischer1979electrical}. In a similar vein, the frustrated spin-$\frac{1}{2}$ system embodying a square lattice, Sr$_2$CuTe$_{0.5}$W$_{0.5}$O$_6$, exhibits glassy behavior below 1.7 K. Below the spin-glass temperature, it is evidenced from the time-of-flight neutron scattering measurements that the imaginary part of the dynamic susceptibility proportional to the energy transfer, $\chi^{\prime\prime}(q,\omega) \sim \hslash\omega$, and the Goldstone modes emerge for $\hslash \omega < k_B T_\text{g}$ \cite{PhysRevLett.127.017201}, and the specific heat shows quadratic behavior below $T_g$\cite{watanabe2018valence}, which indicates the linear dispersion of low-lying magnon excitations. NMR shares an interface with neutron scattering, and both experimental techniques track similar spin dynamics in frustrated magnets. 

The imaginary part of the dynamic susceptibility, \(\chi''(\omega)\) obtained from inelastic neutron scattering, offers a crucial distinction HS mode-driven spin jam in frustrated spin glass including Kitaev magnets and canonical spin glass behavior on a complementary scale. This can be expressed as: $\chi''(\omega) \propto \left(\tan^{-1}\left(\frac{\omega}{\Gamma_{\text{min}}}\right) + \alpha \cdot \left[\frac{\omega}{\omega^2 + \Gamma_L^2}\right]\right)$. The first term, represented by the arctangent function, characterizes HS modes that arise due to quantum fluctuations and are associated with a broad distribution of spin relaxation rates, where \(\Gamma_{\text{min}}\) denotes the lower energy scale of dissipation with $\alpha$ a weighting factor~\cite{yang2015spin, hu2021freezing}. Remarkably, these long-wavelength collective excitations dominate in materials near pristine conditions with minimal disorder, such as SrCr\(_{9p}\)Ga\(_{12-9p}\)O\(_{19}\) at high spin densities (\(p > 0.8\))~\cite{yang2015spin}. In this regime, the system exhibits coherent low-energy spin dynamics that harden with disorder, while the quadratic specific heat behavior ($C_m\sim T^2$) reflects the frozen but collective nature of HS-like excitations akin to Goldstone modes. In contrast, the Lorentzian term \(\frac{\omega}{\omega^2 + \Gamma_L^2}\) reflects localized glassy dynamics within short-range correlated spin clusters, characteristic of canonical spin glasses, where \(\Gamma_L\) sets the characteristic energy at which the susceptibility $\chi^{\prime\prime}$ peaks~\cite{kofu2021spin, PhysRevResearch.6.013006}. As disorder increases, for instance, through nonmagnetic doping in SrCr\(_{9p}\)Ga\(_{12-9p}\)O\(_{19}\), the Lorentzian contribution becomes increasingly significant, signaling a crossover to a conventional spin-glass state dominated by localized excitations~\cite{yang2015spin}. The competition between spin jam and conventional spin glass highlights the delicate interplay between quantum fluctuations and disorder in frustrated systems. At high spin densities, the dynamics remain HS-dominated with long-range coherence, while increased disorder facilitates a gradual transition to a localized, glassy regime~\cite{yang2015spin, syzranov2022eminuscent}. This physical insight not only clarifies the evolution of spin dynamics in systems under doping and disorder but also provides a unifying framework for understanding emergent behaviors in frustrated magnets.

Elastic neutron scattering data of Kitaev
magnet Na\(_2\)Ir\(_{0.89}\)Ti\(_{0.11}\)O\(_3\) reveals profound insights into the underlying physics of topological spin freezing. The wave vector dependence of magnetic structure factor below the spin-glass temperature highlights the emergence of prominent magnetic correlations at low temperatures, indicative of freezing into a topologically nontrivial state~\cite{samarakoon2017scaling}. This is evidenced by the elastic scattering intensity peak centered at the wave vector \(Q = 0.87 \, \text{\AA}^{-1}\), corresponding to short-range spin correlations with a characteristic length scale \(\xi\), given by \(\xi = \frac{1}{\text{FWHM}}\)~\cite{boothroyd2020principles}, where the full width at half maximum (FWHM) was obtained from the digitized graph from the original literature and is found to be \( \text{FWHM} = 0.3 \, \text{Å}^{-1} \) and yielding \(\xi \approx 3.3 \, \text{\AA}\)~\cite{samarakoon2017scaling}. A relatively low value of spin wave velocity, ($c=\frac{\xi k_BT}{\hslash}\sim 65$ m/s)~\cite{PhysRevB.39.2344} at 1.5 K, suggests the low-energy excitations are localized, and the spin correlations decay over short distances. These results underscore the unique magnetic structure factor of this Kitaev magnet, starkly contrasting with the featureless profile of conventional spin glasses like Cu$_{1-x}$Mn with low Mn(2\%) concentration~\cite{samarakoon2017scaling, PhysRevResearch.6.013006}. The neutron scattering results of Cu$_{1-x}$Mn({2\%}) reveal a broad, featureless magnetic structure factor indicative of long-range Ruderman-Kittel-Kasuya-Yosida (RKKY) type disordered spin interactions between dilute magnetic impurities. However, with increasing Mn concentration in Cu$_{1-x}$Mn and in other canonical spin glasses, inelastic neutron scattering reveals a broad low-energy peak, identified as the magnetic boson peak~\cite{kofu2021spin, PhysRevResearch.6.013006, PhysRevB.81.060410}. The hierarchical energy landscape of spin glasses leads to localized excitations within the metastable states~\cite{villain1980dynamics, binder1986spin}, where larger spin clusters generate low-energy modes, and smaller clusters produce higher-energy excitations, exhibiting Bose scaling and a broad spectrum with a high-energy tail~\cite{PhysRevResearch.6.013006}. An intriguing direction for future investigation lies in uncovering a potential connection between the emergence of the magnetic boson peak in canonical spin glasses and the emergence of spin wave modes in topological spin freezing, which may reveal an underlying transition or unifying framework connecting these distinct phenomena.

This corroborates the thermodynamic, NMR results,  and subsequent analysis presented here. A detailed comparison of NMR and neutron scattering results in frustrated magnets may provide profound insights in this context, which is beyond the scope of the present study. Nonetheless, the current brief elucidation of dynamic spin susceptibility in a representative Kitaev magnet offers an exciting impetus for future research directions in frustrated Kitaev magnets.

\section*{Effect of doping on spin freezing}
The influence of external perturbations, such as chemical pressure and magnetic field, is quite striking and often drives quantum phase transitions such as Bose-Einstein condensation, spin liquids,  superconductivity, and non-trivial spin-glass in frustrated magnets. In canonical spin glass systems, the introduction of dopants has been observed to evoke a pronounced and significant influence on the underlying physical mechanism at play. It is known that minute amount of magnetic dopants such as Mn and Fe into non-magnetic host materials, such as Cu$_{1-x}$Mn$_x$, Au$_{1-x}$Fe$_x$, and Ag$_{1-x}$Mn$_x$, there is a substantial increment in the spin-glass temperature upon increasing the doping concentration. This enhancement of $T_g$ is characterized by a marked and significant scaling response that is contingent upon the concentration of dopants, as depicted in Fig. 5a \cite{da2019spin}. In frustrated magnets, the spin-glass temperature in a spin jam state depends on both the energy of barrier mode ($E_B$) necessary for the collective spin flipping and the spin correlation length $\xi (T)$, which can be expressed as $T_g \propto F(E_{B}, \xi)$~\cite{klich2014glassiness}. The spin-glass temperature $T_g$ remains unchanged upon substitution of non-magnetic impurities in frustrated magnets such as SCGO and BCGO until the separation between non-magnetic impurities becomes comparable to $\xi$ \cite{martinez1992magnetic, bono2004intrinsic}.

The effect of doping was investigated in Kitaev materials Ru$_{1-x}$Cr$_x$Cl$_3$, Na$_2$Ir$_{1-x}$Ru$_x$O$_3$ and Na$_2$Ir$_{1-x}$Ti$_x$O$_3$ decorated on a honeycomb lattice, and by varying the doping parameter $x$, it was observed that within a percolation threshold, the doping is less robust compared to the canonical spin glass \cite{mehlawat2015fragile, manni2014effect}. The emergence of topological quantum state akin to Kitaev  QSL in the presence of external magnetic field in the celebrated honeycomb lattice $\alpha\text{-}$RuCl$_3$ is quite remarkable~\cite{do2017majorana}. The well studied Kitaev magnet Na$_2$IrO$_3$ shows a zig-zag antiferromagnetically ordered state at temperature $T_N= 15 $ K~\cite{singh2010antiferromagnetic} and a few percentages of Ru$^{4+}$ doping at Ir$^{4+}$ site gives rise to a topologically protected spin-glass state~\cite{mehlawat2015fragile}. In the context of Kitaev materials, topological freezing refers to the emergence of a glassy state from an underlying topologically ordered phase, such as the Kitaev spin liquid. The term "topological" in this case is not merely a descriptor of the spin freezing but reflects the fact that the freezing emerges from a substrate with inherent topological properties, such as non-trivial ground-state degeneracy and fractionalized excitations that proposed in Kitaev spin liquid~\cite{villain1980dynamics, sen2015topological}. The substitution of Ru$^{4+}$ at Ir$^{4+}$ site preserves the magnetic exchange path that suggests the emergence of glassiness can be intrinsic. This could manifest as a spin jam state, given its correlation with intrinsic glassy states.~\cite{klich2014glassiness}. As presented in Fig. 5a, the spin-glass temperature does not change drastically for the Ru$^{4+}$ doping concentration $0.1\leq x\leq0.3$ in the single crystal of Na$_2$Ir$_{1-x}$Ru$_x$O$_3$. Upon increasing the amount of Ru$^{4+}$ doping concentrations at Ir$^{4+}$ site in Na$_2$Ir$_{1-x}$Ru$_x$O$_3$ and the non-magnetic Ti$^{4+}$ at Ir$^{4+}$ site in Na$_2$Ir$_{1-x}$Ti$_x$O$_3$, the spin-glass temperature decreases smoothly. The observed behavior of spin-freezing temperature upon doping concentration is consistent with spin jam theory~\cite{samarakoon2017scaling, PhysRevLett.127.017201}. Interestingly, upon increasing doping concentration, we observe an increase in $\chi(T_g)$ (see Fig. 2c), while the $T_g$ value is suppressed which, in turn, imposes the constraint $d\chi(T_g)/dT < 0$. 
The observed trend differs from the conventional spin glasses, wherein an escalation in doping concentration typically yields an increase in both $\chi(T_g)$ and $T_g$. This behavior adheres to the underlying relationship expressed by $d\chi(T_g)/dT > 0$ in the case of conventional spin glass~\cite{syzranov2022eminuscent}. It  is worth noting that the extensively studied Kitaev material $\alpha$-RuCl$_3$ hosts a glassy phase upon substitution of $S = 3/2$ Cr$^{3+}$ ions at $J\textsubscript{eff}=1/2$ Ru$^{4+}$ site and the glassy phase remains stable for values of $x$ exceeding 0.1 in $\alpha$\text{-}Ru$_{1-x}$Cr$_x$Cl$_3$ \cite{bastien2019spin}. In contrast, nonmagnetic doping of Ir$^{3+}$ ($5d^6$: $J_\text{eff}=0$) in $\alpha$\text{-}Ru$_{1-x}$Ir$_x$Cl$_3$ stabilizes a quantum disordered QSL like state for $0.16\leq x\leq0.4$~\cite{lampen2017destabilization}. The data collapse behavior in specific heat is observed as a function of the magnetic field, and it is achieved by plotting $CH^{1-\gamma}/T$ vs $T/H$, where $\gamma$ (= 0.19) represents the critical exponent typical for random singlets in the quantum disordered state ~\cite{do2020randomly}. The observed features suggest that frustrated Kitaev magnets are quite sensitive to external stimuli in undergoing a phase transition leading to the emergence of a plethora of topological states driven by perturbations in a controlled manner. Taking into account the intriguing phenomena detected in  Kitaev magnets owing to a subtle interplay between competing degrees of freedom, their free energy landscape and effect of external perturbations on the underlying physical phenomena, we proposed a comprehensive phase diagram as shown in Fig. 5b that reflects immense promise for the experimental realization of topological and quantum states such as Kitaev QSL with elusive Majorana fermions in this class of frustrated quantum materials.
\begin{figure*}[ht]
\centering
\includegraphics[width=\linewidth]{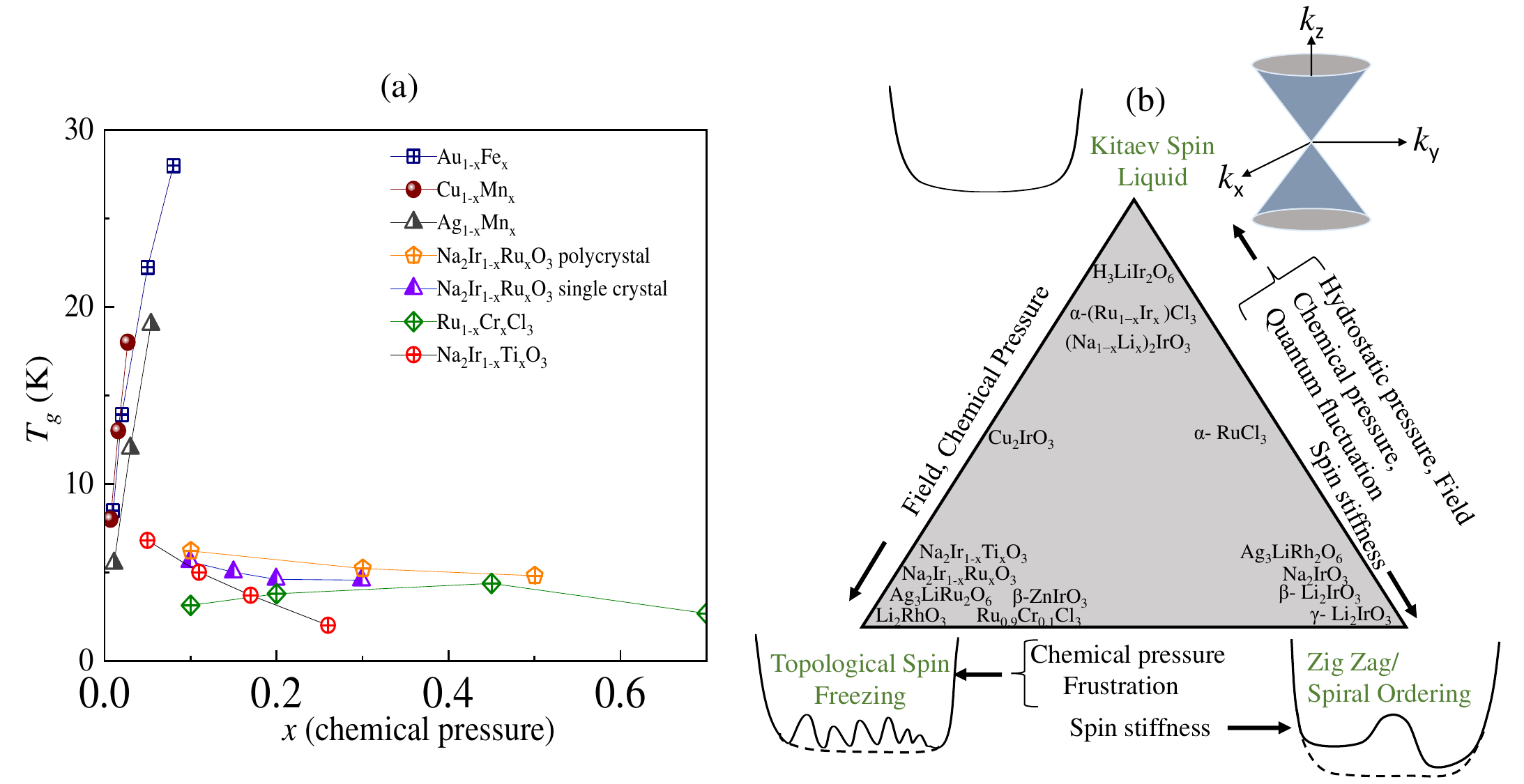}
\caption{(a) The spin-glass temperature $T_g$ as a function of doping concentration $x$ for canonical spin glass materials AuFe, CuMn, and AgMn, as well as for frustrated Kitaev spin glasses adapted from references as indicated~\cite{da2019spin, mehlawat2015fragile, bastien2019spin, manni2014effect} (b) Schematic triangular phase diagram of selected Kitaev materials, distinguishing unique characteristics based on their low-energy, non-trivial excitations and energy landscapes. Zig-zag and spiral ordering are associated with well-defined energy minima indicative of classical magnetic order. Topological spin freezing is characterized by a rugged energy landscape, reflecting glassy dynamics arising from frustration and disorder effects. The Kitaev spin liquid phase accommodates both Dirac cone and flat band dispersions, which emerge under different conditions as signatures of the spin liquid state. (see~\cite{lee2023coexistence, li2018role, do2020randomly, cao2013evolution, abramchuk2017cu2iro3, choi2019exotic, bahrami2022first, liu2011long, takayama2015hyperhoneycomb, kimchi2015unified, khuntia2017local, mehlawat2015fragile, manni2014effect, bastien2019spin, haraguchi2022quantum, zubtsovskii2022topotactic, haraguchi2023monoclinic} for more details).}
\label{fig}
\end{figure*}

\section*{A Proposed Phase Diagram for Kitaev Materials}
To construct a phase diagram for Kitaev magnets, which captures transitions between complex magnetic ordering, topological spin freezing, and Kitaev spin liquid like behavior, we account for three primary parameters: applied magnetic field, hydrostatic pressure, and chemical pressure. These external factors allow us to tune the interactions and probe different phases of the material. The general Hamiltonian governing a Kitaev model is given by:
$\mathcal{H} = \sum_{\langle ij \rangle} \left( -K_{\gamma} S_{i}^{\gamma} S_{j}^{\gamma} + J \mathbf{S_i} \cdot \mathbf{S_j} + \Gamma \left( S_i^{\alpha} S_j^{\beta} + S_i^{\beta} S_j^{\alpha} \right) \right)$, where \( K_{\gamma} \) represents the anisotropic Kitaev exchange along different bond directions \( \gamma \), \( J \) is the Heisenberg exchange interaction, and \( \Gamma \) denotes the off-diagonal symmetric exchange involving spin components \( \alpha \) and \( \beta \)~\cite{takagi2019concept}. These interactions are crucial for describing the intricate balance between quantum spin liquid and magnetically ordered phases, as well as topological spin freezing, influenced by external perturbations. In real Kitaev materials, such as $\alpha$-RuCl$_3$~\cite{sears2015magnetic} and $\gamma$-Li$_2$IrO$_3$~\cite{biffin2014noncoplanar}, the ideal quantum spin liquid ground state predicted by the pure Kitaev model is disrupted by additional perturbations, such as Heisenberg exchange, off-diagonal symmetric exchange, and further-neighbor interactions. These deviations from the pure Kitaev model give rise to long-range magnetic order at low temperatures such as zigzag or spiral patterns. The intricate balance between these interactions not only stabilizes these ordered phases but also offers a deeper understanding of the glassy behaviors and emergent phases in Kitaev magnets.

The emergence of glassiness in Kitaev materials, particularly in systems like \( \alpha \)-RuCl$_3$, arises from the interplay between high \( Z_2 \) vortex density, magnetic fields, and the intrinsic topological constraints of the spin-liquid phase~\cite{yogendra2023emergent1, yogendra2024fractional}. As the magnetic field exceeds a critical value, the system transitions into a dense vortex regime where the proliferation of \( Z_2 \) fluxes generates local frustration, impeding long-range order and trapping the system in numerous metastable states~\cite{yogendra2024fractional, takahashi2024z_2}. These flux excitations, characterized by the flux operator \( W_p \), exhibit slow dynamics that mirror the energy landscape of classical spin glasses but are distinctly topological. This emergent freezing phase retains quasi-long-range correlations with a substantial flux correlation length, despite the absence of crystallization~\cite{yogendra2023emergent1}. The autocorrelation function \( \langle W_p(t)W_p(0) \rangle \), which decays slowly, reflects the sluggish dynamics of the glassy phase. The spin-lattice relaxation rate \( \frac{1}{T_1} \) is related to the fluctuating local magnetic field \( h_{\text{loc}}(t) \) of electronic origin at the nuclear site, coupled to the nuclear spins as:$
\frac{1}{T_1} \propto \int_0^\infty \langle h_{\text{loc}}(t)h_{\text{loc}}(0) \rangle \, dt,
$where \( \langle h_{\text{loc}}(t)h_{\text{loc}}(0) \rangle \) is the autocorrelation function of the local magnetic field~\cite{khuntia2017local}. In a conventional magnetic material, this correlation decays rapidly as spin fluctuations die out. However, in this emergent frozen state, the decay slows significantly due to the quasi-static nature of $Z_2$ fluxes or Majorana modes. Consequently, the NMR signal, rather than freezing completely, will result in a finite relaxation rate down to the lowest experimentally accessible temperatures. This behavior can be probed directly through NMR experiments, providing valuable insight into the glassy dynamics and persistent flux correlations. The enhanced second- and third-order susceptibilities in this phase~\cite{holleis2021anomalous}, along with the slow autocorrelation decay, highlight the non-Gaussian and chaotic nature of the fluctuations~\cite{yogendra2023emergent1}. The system becomes trapped in local minima, forming a glassy phase dominated by flux operators rather than spins, making Kitaev magnets a unique platform to explore glassy behavior driven by quantum fluctuations, non-trivial excitation spectra,  and short-range spin correlations.

The substitution of Ag (which is larger and less electronegative than Li) tends to expand the lattice due to the increased ionic radius leading to the  decrease of chemical pressure in going from Li$_2$RhO$_3$ to Ag$_3$LiRh$_2$O$_6$. The transition from Li$_2$RhO$_3$ to Ag$_3$LiRh$_2$O$_6$ involves a topochemical reaction that introduces Ag ions, resulting in a trigonal distortion of the octahedral coordination around Rh~\cite{bahrami2022first}. This structural modification enhances the competition between spin-orbit coupling and crystal field effects, leading to the emergence of a new magnetic ground state characterized by robust antiferromagnetic ordering. Consequently, Ag$_3$LiRh$_2$O$_6$ exhibits significantly stronger magnetic interactions compared to the spin-glass behavior observed in Li$_2$RhO$_3$~\cite{bahrami2022first}. The intricate interplay between quantum fluctuations, spin rigidity, and exchange frustration in magnetic systems gives rise to a rich tapestry of emergent phases, elegantly captured in a triangular phase diagram (Fig. 5b). The vertices of this diagram represent distinct physical states: the Kitaev spin liquid (KSL), characterized by fractionalized excitations and long-range quantum entanglement; topological freezing, where disorder and frustration arrest spin dynamics into a glassy, localized state; and zig-zag or spiral ordering, defined by classical long-range magnetic order. The edges connecting these phases are governed by parameters such as strength of quantum fluctuation(Q), spin stiffness, chemical pressure, hydrostatic pressure, and exchange frustration. Understanding these transitions not only deepens our knowledge of magnetic frustration but also offers insights into topological phases, spin dynamics, and disorder effects. Spin stiffness representing the rigidity of the spin configurations, is defined as: $
\rho = \frac{1}{N} \sum_q \left( \frac{\partial^2 E(q)}{\partial q^2} \right),$ where $E(q)$ is the spin wave dispersion. Systems with high stiffness stabilize classical long-range orders, such as zig-zag or spiral configurations, whereas reduced stiffness enables localized or frozen states. Quantum fluctuations, arising from zero-point motion, renormalize this stiffness through corrections (\(\Delta \rho(Q)\)), yielding an effective stiffness $
\rho_{\text{eff}} = \rho_0 - \Delta \rho(Q)$~\cite{PhysRevLett.60.1057, PhysRev.86.694}. Physically, strong quantum fluctuations (\(\Delta \rho(Q) \sim \rho_0\)) in the Kitaev spin liquid phase collapse \(\rho_{\text{eff}}\), leading to a dynamic, entangled state without classical long-range order. In the topological freezing phase, intermediate stiffness (\(\rho_{\text{eff}} > 0\)) reflects a balance between frustration and fluctuations, stabilizing localized, non-coherent spin dynamics~\cite{halperin1977hydrodynamic}. Conversely, in the zig-zag or spiral ordering phase, high stiffness (\(\rho_{\text{eff}} \approx \rho_0\)) enables robust classical order, with quantum fluctuations playing only a perturbative role~\cite{halperin1969hydrodynamic}. This interplay between spin stiffness and quantum fluctuations underscores the emergence of diverse magnetic phases and their transitions in Kitaev systems, highlighting the intricate coupling of quantum and classical effects.

In the phase diagram (Fig. 5b inset), we highlight that Kitaev magnets can exhibit two distinct low-energy features: a Dirac band or a flat band, depending on the specific parameter regime of the Hamiltonian. This duality arises from the delicate balance between the gauge symmetries and energy dispersion within the Kitaev model~\cite{PhysRevB.78.115116, rousochatzakis2018quantum, takagi2019concept, yogendra2024fractional}. In the uniform flux configuration (i.e., \( W_p = +1 \) for all plaquettes), the low-energy spectrum exhibits gapless Dirac points, reminiscent of graphene's Dirac cone structure. However, when next-nearest-neighbor interactions, represented by the term proportional to \( K \), are introduced, they break time-reversal symmetry, thereby gapping out the Dirac points and leading to topologically non-trivial Chern bands characterized by a quantized Chern number~\cite{takagi2019concept, yogendra2024fractional}. In specific gauge configurations, the energy dispersion flattens, leading to flat bands. The emergence of flat bands in Kitaev magnets originates from the cancellation of kinetic energy contributions of Majorana fermions, driven by the destructive interference of hopping matrix elements. The hopping matrix \( T_{ij} \), which governs the amplitude and phase of fermion hopping between sites \( i \) and \( j \), is intrinsically coupled to the background \( \mathbb{Z}_2 \) gauge fields~\cite{yogendra2024fractional}. Decomposing \( T_{ij} \) into symmetric and antisymmetric components reveals their distinct roles in shaping the energy dispersion. When the antisymmetric component dominates, the interference of hopping phases leads to a suppression of the effective kinetic energy, flattening the band structure over specific regions of the Brillouin zone and resulting in highly degenerate energy states. This can be captured by the flattening of the energy dispersion \( \epsilon(k) \) over the Brillouin zone, where the second derivative \( \partial^2 \epsilon(k) / \partial k^2 \approx 0 \), indicating negligible curvature and thus a flat band~\cite{yogendra2024fractional}. In the classical limit, the flat energy landscape is associated with to the degeneracy of ground states in the large-spin (\( S \)) Kitaev model. The total energy for each dimer configuration, which minimizes the energy, scales as \( -JSN/2 \), and the number of these configurations grows exponentially with the number of lattice sites \( N \), scaling as \( (1.662)^N \)~\cite{PhysRevB.78.115116}. These degenerate ground states are discrete in spin space but are connected by continuous "valleys" in energy, producing a flat band-like behavior. Even when quantum fluctuations are included via a spin-wave analysis (which introduces corrections in powers of \( 1/S \)), the system retains this flatness, reflecting the resilience of the classical degeneracy to quantum corrections~\cite{PhysRevB.78.115116, rousochatzakis2018quantum}. Hence, the flat band observed in both the quantum and classical regimes of the Kitaev model reflects deep connections between gauge symmetries, frustration, and the system's topological structure. The experimental observation of flat-band energy dispersion in the promising $J_\text{eff} = 1/2$ Kitaev magnet Na$_2$Co$_2$TeO$_6$ is a step forward in this direction~\cite{PhysRevLett.110.097204, PhysRevLett.129.147202, zhang2024microbial}. The existence of flat-band energy dispersion is supported by the prevalence of significant ground state degeneracy, short - range spin correlations, and gapless excitation spectra, as probed by thermodynamics, NMR, and INS experiments on the Kitaev materials presented here. This scenario elucidates the innate connection between topology, electron correlation, and symmetry, signaling the rich potential of this class of frustrated magnets in realizing novel organizing principles and quantum phenomena~\cite{wen2017colloquium, kitaev2006anyons, khatua2023experimental, takagi2019concept}.

\section*{Discussion}

The honeycomb magnet stabilizes on a bipartite lattice, and the Heisenberg exchange interaction between spins leads to a N\'eel state at low temperature. However, spin-orbit driven anisotropic bond-dependent frustrated Kitaev exchange interaction between $J_\text{eff} = 1/2$ spins decorated on a honeycomb lattice can host a quantum spin liquid state with deconfined fractional excitations~\cite{kitaev2006anyons, khatua2023experimental}. The frustrated Kitaev magnets on a honeycomb lattice are potential candidates to host non-trivial spin-freezing of topological origin~\cite{yogendra2023emergent1}. 
The order function formalism associated with remanent properties of frustrated magnets may be of paramount importance in elucidating a broad class of disordered systems including spin glass~\cite{fukuyama1982anderson}.
The low-temperature spin freezing behavior observed in pristine and doped frustrated Kitaev magnets Li$_2$RhO$_3$, Na$_2$Ir$_{1-x}$Ti$_x$O$_3$, Ru$_{1-x}$Cr$_x$Cl$_3$, Li$_2$Ir$_{1-x}$Ti$_x$O$_3$, and Na$_2$Ir$_{1-x}$Ru$_x$O$_3$ exhibit characteristic features including low-energy excitations akin to the hydrodynamic modes with gapless spectra proposed for frustrated spin glass materials~\cite{halperin1977hydrodynamic}. The presence of a linearly dispersing mode is evident in the low-temperature thermodynamic, NMR, and neutron scattering experiments. The $T^2$ behavior of magnetic specific heat below the freezing temperature remains independent of the magnetic field, which is quite different from that observed in conventional spin-glass materials. The higher value of the characteristic energy scale, $\tau_2$, with respect to the freezing temperature $T_g$ (Table 1), can be due to the underestimation of next-nearest neighbor interactions in the HS framework; however, this can be addressed by increasing the number of nearest neighbor $z$ in honeycomb spin lattice from three to some effective higher number. 

In general, the conventional spin glass materials do not support the spin wave excitations where the spin stiffness $\rho_s=0$.  Spin wave excitations arise from clusters of spins, each of which exhibits finite bounds with finite spin stiffness constant. These spin clusters undergo independent spin freezing that is attributed to jamming effects~\cite{klich2014glassiness}. The observed spin fluctuations in frustrated Kitaev magnets, as probed by neutron scattering
measurements, extended down to freezing temperatures, likely originating from the spins located at the boundaries of spin clusters~\cite{li2023evolution}. Taking into account Anderson's proposal regarding the areal scaling of free energy fluctuations at the boundaries of clusters, one can expect the emergence of a rugged energy landscape beneath an initially flat energy landscape~\cite{anderson1978concept, chandra1993anisotropic}. It is worth noting that
the HS mode is coupled with both electronic and nuclear magnetic moments due to hyperfine coupling between the nucleus and the surrounding electronic moments~\cite{podolsky2009halperin}. Understanding the nature of spin correlations and fluctuation of hyperfine fields in the ground state of the frustrated Kitaev magnet is highly significant, which may underpin elusive Kitaev QSL and associated low energy excitations. In this context, microscopic techniques such as NMR provide an excellent probe for tracking the intrinsic magnetic susceptibility,  nature of spin correlations, and associated unconventional low-energy excitations in the ground state of these frustrated magnets~\cite{khatua2023experimental, khuntia2017local}. The power law behavior of the NMR spin-lattice relaxation rate supports thermodynamic results in these Kitaev magnets.

Unlike the hierarchical organization of free energy observed in conventional spin glasses, a non-hierarchical structure is posited, supported by evidence of memory and aging effects in frustrated kagome structures such as SCGO, BCGO, and Kitaev magnets like Li$_2$RhO$_3$ and Na$_2$Ir$_{0.89}$Ti$_{0.11}$O$_3$~\cite{samarakoon2016aging, samarakoon2017scaling}.
The aging behavior exhibits diminished prominence, necessitating a large time scale observations to manifest it. The presence of a subtle memory effect implies the development of a nonhierarchical landscape featuring a wide, nearly flat, but rough bottom~\cite{samarakoon2016aging, yang2015spin}. The non-Abelian nature of defect propagation in anisotropic kagome was proposed for the memory effect in SCGO like nearly defect-free kagome materials \cite{chandra1993anisotropic}. The defect-free hyperhoneycomb $\beta\text{-}$ZnIrO$_3$ and $\beta\text{-}$MgIrO$_3$ Kitaev magnets with dominant ferromagnetic exchange interaction exhibit intrinsic glassiness, necessitating further theoretical investigation.
The HS framework sheds very interesting insights into the spin correlations and spin-freezing mechanism in 2D and 3D  Kitaev magnets. The parameters derived from the analysis of magnetic susceptibility and specific heat results following HS framework on a few selected Kitaev magnets are documented in Table 1. The low-energy excitations of the spin jam state are primarily governed by the HS mode, wherein spin wave excitations emerge from the unconventional spin-freezing dynamics. This is in stark contrast to conventional spin glasses characterized by localized two-level energy systems, where the excitations occur through the barrier mode.
Our analysis suggests that the HS  model is quite apt and generic in elucidating the low temperature spin freezing mechanism in a large class of frustrated quantum magnets~\cite{halperin1977hydrodynamic}.

Overall, while the semiclassical framework of the HS formalism and spin-jam theory may initially seem incompatible with the extreme quantum spins of Kitaev magnets studied in this paper, the experimental evidence highlights the need for further theoretical development to fully elucidate the underlying physics. In the quantum regime, quantum fluctuations due to frustration can lead to a multitude of metastable states and quantum tunneling effects, differing from the semiclassical picture. This complexity is illustrated in the spin-1/2 Heisenberg-Kitaev kagome magnet, where anisotropic Kitaev interactions drive the quantum-to-classical crossover~\cite{yang2020quantum}. The spin-1/2 system, Sr$_2$CuTe$_{0.5}$W$_{0.5}$O$_6$ exhibits glassy behavior below $T_g \approx 1.7$ K. The imaginary part of dynamic susceptibility $\chi^{\prime\prime} \propto \hslash \omega$ for $\hslash \omega < k_B T_g$, suggesting low-energy Goldstone modes~\cite{PhysRevLett.127.017201}. The specific heat shows a quadratic temperature dependence, $C \propto T^2$, below the spin-freezing temperature, indicating linear dispersion of low-lying magnon excitations~\cite{watanabe2018valence}. Additionally, the specific heat exhibits a broad maximum around 1.7 K., with field-independent behavior up to 9 T. These experimental findings suggest the presence of a quantum analog of spin jam in this diluted and low-spin quantum material, which could provide deeper insights into the nature of disordered quantum magnets and their excitations. The randomness leaves the emergent gauge fields as distinct degrees of freedom and prevents the system from undergoing order by disorder, leading to the formation of a frozen state with spontaneously broken symmetry under global spin rotations~\cite{PhysRevLett.127.017201}. This study provides insights into how weak exchange randomness influences the emergence of hydrodynamic modes in low-spin frustrated magnets~\cite{PhysRevB.101.024413}, shedding light on the intricate interplay between emergent gauge fields and low-energy excitations in these systems. The rare-earth triangular oxide TmMgGaO$_4$ features Tm$^{3+}$ ions with a 4$f^{12}$ electron configuration and a spin–orbit moment \( J = 6 \) with strong Ising anisotropy~\cite{hu2020evidence}. Crystal-electric-field splitting separates these ions into a non-Kramers doublet, well sepaarated from other energy levels by about 400 K, allowing the doublet to be treated as an effective spin-1/2 (\( J_{\text{eff}} = 1/2 \)) system \cite{hu2020evidence, li2020partial}. In this quantum Ising magnet, the presence of quenched disorders, specifically weak non-magnetic disorders due to Ga and Mg site sharing, gives rise to the Halperin-Saslow mode. It is proposed that weak quenched disorder transforms the emergent 2D Berezinskii–Kosterlitz–Thouless (BKT) phase and the critical region into a gauge glass \cite{huang2024emergent}. This transformation leads to the emergence of the Halperin-Saslow mode associated with the gauge glass in the quantum Ising magnet TmMgGaO$_4$. In applying a hydrodynamic approach to model the disordered Kitaev materials, we recognize that this method serves as a phenomenological tool rather than a microscopic theory. However, the hydrodynamic framework effectively captures the essential low-temperature behaviors, including gapless nature of low-energy excitations that emerge in the presence of disorder. This approach allows us to describe the collective modes and associated thermodynamic responses that arise from the complex interplay of spin-orbit coupling, Mott physics, and disorder without requiring detailed microstate calculations. As such, our model represents a simplified yet scientifically grounded interpretation of the magnetic dynamics observed in these materials.

The possible scenarios that can elucidate the observed unconventional low temperature behavior of frustrated Kitaev magnets are: (i) In the frame of the Kitaev model,  the distinctive spin-orbit driven bond-dependent anisotropic interactions engender a set of conserved quantities intimately associated with the spins, often denoted as integrals of motion~\cite{takagi2019concept}. The existence of this substantial ensemble of conserved quantities such as $Z_2$ flux operator $W$ can facilitate the emergence of weakly broken ergodicity, thereby leading to the intrinsic manifestation of glassy behavior in the proximity of a clean system without doping~\cite{serbyn2021quantum}. (ii) In the absence of ZFC-FC bifurcation, the analysis of higher-order susceptibility can provide valuable insights into the intrinsic spin-glass phase, manifesting distinctive anomalies in the vicinity of the glass transition. (iii) Remarkably, intrinsic glassiness represents a plausible magnetic phase that can endure within the domain of quantum fluctuation and topological defects that may underpin elusive Kitaev QSL.
\section*{Conclusion}
Frustrated Kitaev quantum materials are potential candidates to host intriguing quantum phenomena such as topological spin glass and non-trivial low-energy excitations, understanding of which may reveal vital clues to the sought-after Kitaev QSL. Our comprehensive analysis and discussion, based on Halperin-Saslow hydrodynamic linearly dispersing modes, elegantly capture the essence of the spin-freezing mechanism, with spin jam providing results consistent with the Halperin-Saslow theory. The power law behavior of magnetic specific heat at low temperature in a few representative Kitaev magnets decorated on 2D and 3D spin lattices manifests the persistence of abundant low-energy topological excitations in conjunction with macroscopic ground state degeneracy, slow spin dynamics, and short range spin correlations which is in sharp contrast to that observed in canonical spin-glass magnets. In addition, our analysis of magnetic specific heat below the spin-freezing temperature in Kitaev spin glass materials following the HS framework suggests the dominant role of low-energy excitations. Our results based on the comparative account of thermodynamic NMR, and neutron scattering experiments in a complementary scale demonstrate that frustrated Kitaev materials are promising candidates to exhibit an intrinsic glassiness, the genesis of which can be attributed to the effect of quantum fluctuations, low energy excitations, and the imposition of topological constraints. The distinctive character of the free energy landscape and non-trivial behavior of frustrated Kitaev magnets enable us to establish a rich phase diagram that may provide a new direction for further exploration of 2D and 3D Kitaev magnets' are potential contenders to host quantum and topological states with elusive fractional excitations. The measurement of non-linear susceptibility as a direct probe of the Edward-Anderson order parameter may provide deep insights into the free energy landscape and non-trivial low temperature spin-freezing in frustrated Kitaev magnets. A detailed inelastic neutron scattering and low-temperature NMR experiments on the high quality single crystals conducted below the spin-glass temperature may provide profound insights into the low-energy excitations in this class of frustrated magnets. Such insights are pivotal for understanding topological order and for detecting and manipulating exotic fractional excitations in Kitaev magnets. The intriguing question of whether frustrated Kitaev magnets can exhibit spin-freezing without quenched disorder prompts further investigation. Additionally, the prevalence of spin-glass in proximate Kitaev quantum spin liquid candidates invokes further studies in this direction. Such efforts may broaden our understanding of complex magnetic systems, establish realistic Hamiltonians, and pave the way to gain new insights into the behavior of frustrated magnets under the influence of various external perturbations such as doping, pressure, and applied magnetic fields. These are crucial for underpinning novel topological and quantum states and associated low-energy excitations, as well as the organizing principles for designing Kitaev quantum materials with optimal functionalities.

The phenomenological scenario proposed in this work is simple, yet it captures some of the key attributes of macroscopic ground state degeneracy, low-energy excitations, and short-range correlations, as manifested in complementary thermodynamic, NMR, and neutron scattering experiments in emblematic frustrated Kitaev magnets that host non-trivial spin-freezing phenomena. This invokes comprehensive microscopic theoretical models for a deeper understanding of promising frustrated magnets that may eventually lead to the experimental realization of novel quantum phenomena with exotic quasi-particle excitations, essential for addressing some of the fundamental questions in condensed matter and a step forward for next-generation technologies to meet the challenges of ever-increasing demands for energy storage and computing.
\section*{Method}
\subsection*{$\tau_1$ and $\tau_2$ for doped Kitaev magnets}
In the presence of doping with two magnetic elements of the same site, the two temperature scales can be expressed as:
\begin{equation}
\begin{aligned}
    E_1 =&\frac{2(xg_2\mu_B)^2S_2(S_2+1)N_A}{z\chi_{M}} + \\
   &\quad  \frac{2((1-x)g_1\mu_B)^2S_1(S_1+1)N_A}{z\chi_{M}} = k_B\tau_1
    \end{aligned}
\end{equation}
and
\begin{equation}
    E_2 = \frac{3n_p\zeta(3)}{\pi}\frac{k_B^3}{(xg_2\mu_B + (1-x)g_1\mu_B)^2}\frac{\chi_{M}}{C_{m}/T^2} = k_B\tau_2.
\end{equation}
\section*{Data availability}
The Data used to generate Figs. 2, 3, 4, and Table 1 are available from the cited articles
published by the American Physical Society.
\bibliography{KSG2}

\section*{Acknowledgements}
U.J. acknowledges support from the Council of Scientific and Industrial Research, India. P.K. acknowledges funding from the Science and Engineering Research Board and Department of Science and
Technology, India through research grants.

\end{document}